\documentclass[prd,preprint,eqsecnum,nofootinbib,amsmath,amssymb,preprintnumbers,tightenlines]{revtex4}

\usepackage{mathrsfs}
\usepackage{hyperref}
\usepackage{graphicx}
\usepackage{bm}

\def\wn{{\mathfrak w}}
\def\Im{{\rm Im}}
\def\Re{{\rm Re}}
\def\ls2{{\ell_s^2}}
\def\x{{\bf x}}
\def\dd{{\rm d}}
\def\Db{{\mathbb D}}
\def\Dt{{\mathrm D}}
\def\ret{{\rm ret} }

\def\rbar{{\bar r}}
\def\f{{ f}}
\def\hatf{{\hat {\mathcal F}}}
\newcommand\bga{\begin{align}}
\newcommand\nda{\end{align}}
\def\xts{\Delta x_{\scriptscriptstyle \rm TS}}
\def\etav{{\eta} }

\def\x{{\bf x}}
\def\p{{\bf p}}

\def\F{{\mathcal F}}

\def\F{{\mathcal F}}
\def\N{{\mathcal N}}

\def\Re{{\rm Re}}
\def\Im{{\rm Im}}

\def\nc{N_{\rm c}}

\def\st{\begin{equation}}
\def\stp{\end{equation}}
\def\bg{\begin{eqnarray}}
\def\nd{\end{eqnarray}}
\def\Eq#1{Eq.~(\ref{#1})}
\def\app#1{Appendix~\ref{#1}}
\def\Fig#1{Fig.~\ref{#1}}
\def\Sect#1{Section~\ref{#1}}
\def\Ref#1{Ref.~\cite{#1}}
\def\m{{ m}}
\def\T{{T_o}}

\def\llangle{\left\langle}
\def\rrangle{\right\rangle}

\def\N{\mathcal{N}}
\def\w{\mathfrak{w}}

\def\gsim{\mbox{~{\protect\raisebox{0.4ex}{$>$}}\hspace{-1.1em}
	{\protect\raisebox{-0.6ex}{$\sim$}}~}}

\def\nott#1{\setbox0=\hbox{$#1$}                
   \dimen0=\wd0                                 
   \setbox1=\hbox{/} \dimen1=\wd1               
   \ifdim\dimen0>\dimen1                        
      \rlap{\hbox to \dimen0{\hfil/\hfil}}      
      #1                                        
   \else                                        
      \rlap{\hbox to \dimen1{\hfil$#1$\hfil}}   
      /                                         
   \fi}                                         %

\advance\parskip 1.9pt
\advance\voffset -0.2in

\def\st{\begin{equation}}
\def\stp{\end{equation}}
\def\bg{\begin{eqnarray}}
\def\nd{\end{eqnarray}}
\def\nc{{\, ,}}
\def\np{{\, .}}

\begin{document}

\vspace*{1cm}

\preprint{INT-PUB-09-003}
\title{Thermal Noise and Stochastic Strings in AdS/CFT }

\author{Dam T. Son}
\affiliation
   {Institute for Nuclear Theory, 
    University of Washington, 
    Seattle, WA 98195-1550, USA 
 }
\author{Derek Teaney}
\affiliation
    {%
    Department of Physics \& Astronomy,
    SUNY at Stony Brook,
    Stony Brook, New York 11794, USA
    }%

\date{January 2009}



\begin{abstract}
We clarify the structure of  thermal noise in AdS/CFT by studying the dynamics
of an equilibrated heavy quark string. Using the Kruskal extension of the
correspondence to generate the dynamics of the field theory on the Keldysh
contour, we show that the motion of the string is described by the classical
equations of motion with a stochastic boundary condition on the stretched
horizon.  The form of the stochastic boundary condition is consistent with the
dissipation on this surface and is found by integrating out the fluctuations
inside of the stretched horizon.  Solving the equations of motion for the
fluctuating string we determine the full frequency dependence of the random
force on the boundary quark and show that it is consistent with the frequency
dependent dissipation.  We show further that the stochastic motion reproduces
the bulk to bulk two point functions of the Kruskal formalism.  These turn out
to be related to the usual retarded bulk to bulk propagator by KMS relations.
Finally we analyze the stochastic equations and give a bulk picture of the
random boundary force as a flip-flopping trailing string solution.  The basic
formalism can be applied to the fluctuations of gravitons, dilatons, and other
fields.
\end{abstract}
\maketitle

\section{Introduction}

The AdS/CFT correspondence 
 has been used extensively 
to study the properties of strongly coupled gauge theories in a controllable
setting 
\cite{Maldacena:1997re,Gubser:1998bc,Witten:1998qj}.
The correspondence has led to important insights into the
nature of strongly  coupled plasmas.  
In particular, 
Refs. \cite{Policastro:2001yc,Kovtun:2004de,Kovtun:2003wp,Buchel:2003tz,Buchel:2004qq}
 established
that in a large class of gauge theories with gravity duals the 
ratio between the shear viscosity  and the entropy is
\st
  \frac{\eta}{s} = \frac{\hbar}{4\pi} \np 
\stp
This was important because
it showed that there exist certain theories which 
realize
the small transport time scales needed to explain the elliptic
flow observed at RHIC
\cite{Bellwied:2005kq,Adcox:2004mh,Molnar:2001ux,Teaney:2003kp,Romatschke:2007mq,Song:2007fn,Dusling:2007gi,Molnar:2008xj}.
Since this work on shear viscosity many other 
transport properties of strongly 
coupled plasmas have been computed using the
correspondence. Of particular relevance
to this work is the computation of the heavy quark drag  and 
diffusion  coefficient in $\N=4$ Super Yang Mills (SYM) at
large $N_c$ and strong coupling \cite{HKKKY,Jorge,Gubser}.

Most of the time, thermal noise is neglected in AdS/CFT.  
This seems at 
odds with the fluctuation-dissipation theorem  and leads
to some seemingly incorrect results from 
the correspondence. For instance, it predicts the 
absence of long-time hydrodynamic tails \cite{Kovtun:2003vj}, 
zero drag on mesons  
\cite{Chernicoff:2006hi, Liu:2006nn,Peeters:2006iu},
and the lack of Brownian
motion of a quark string in ${\rm AdS}_5$ \cite{HKKKY}. 
In many cases, the effect 
of thermal noise is suppressed either by large $N$ or 
large $\lambda$,  and therefore these inconsistencies with  field 
theory intuition were rationalized as an artifact of these
restrictive limits. 
Certain transport properties
such as meson transport \cite{Dusling:2008tg} and momentum diffusion 
\cite {Jorge,Jorge2,Gubser:2006nz} are intrinsically related to the fluctuations. 
These transport rates were computed using the correspondence 
by computing the drag and appealing to the boundary 
fluctuation-dissipation theorem  to determine the fluctuation rate.
A notable exception  to this rule is the calculation of the 
momentum broadening of a fast  heavy quark 
\cite{Jorge2,Gubser:2006nz}
 which used the Kruskal
formalism to compute this diffusion rate  in
an out of equilibrium setting where the  fluctuation dissipation 
theorem does not apply \cite{Jorge2}. 

The purpose of the present work is to overcome these difficulties
by working through the simplest possible system which should exhibit drag and 
noise in AdS/CFT. This system is the Brownian 
motion of a heavy quark placed in the $\N=4$ SYM plasma. 
A quark in AdS/CFT is represented  as an open string stretching 
from the horizon up to a probe brane. 
A schematic of the AdS geometry together with 
the heavy quark probe and the stretched horizon (see below) 
is shown in \Fig{string1fig}. 
The various problems with large $N_c$ and large $\lambda$ are easily 
summarized by the simple fact that, according to the classical equations
of motion, the string does not move
in the absence of external force. Clearly this is 
not the gravity dual of a Brownian particle in  equilibrium with plasma.
\begin{figure}
\begin{center}
\includegraphics[height=1.5in]{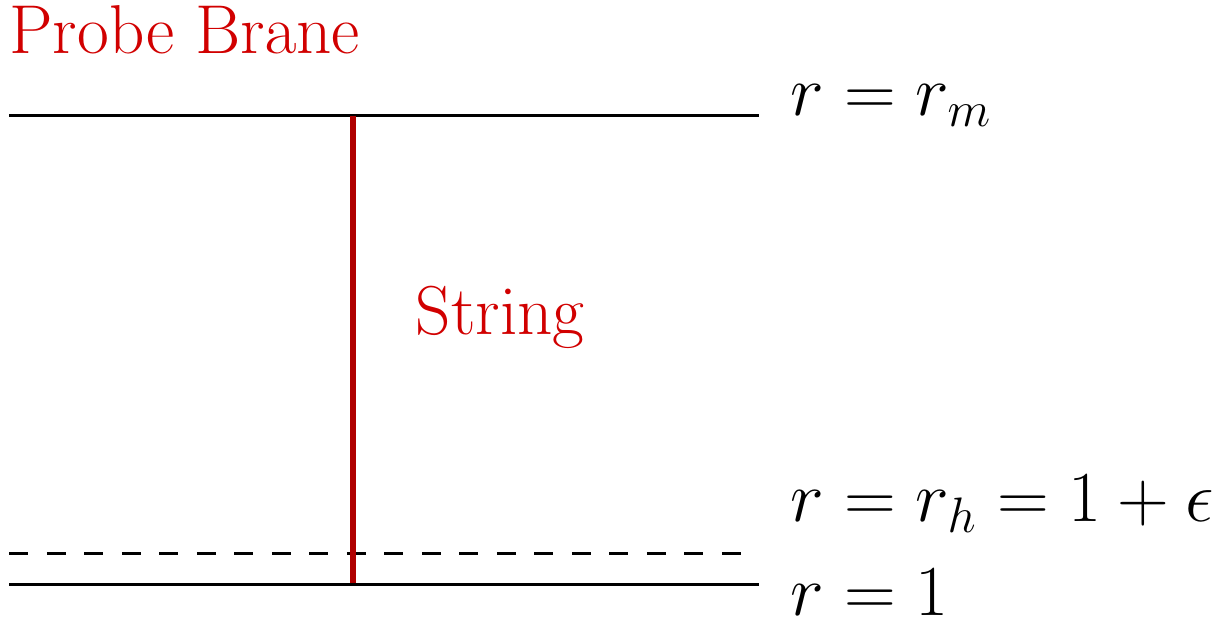} 
\caption{A schematic of a classical string in AdS$_{5}$ corresponding to a 
heavy quark. 
The horizon is at $r=1$ in the coordinates of this work. The stretched
horizon is at $r_h=1+\epsilon$ and the endpoint of the string is at
$r_m$ with $r_m \gg 1$. Gravity pulls downward in this figure. }
\label{string1fig}
\end{center}
\end{figure}

One expects that the black brane should emit Hawking radiation inducing 
random motion on the string \cite{Lawrence:1993sg}. Hawking
radiation is consistent with the fluctuation-dissipation theorem \cite{Hawking:1974sw,Gibbons:1976pt,Gibbons:1976es,Hartle:1976tp,Unruh:1976db,Israel:1976ur},
and this consistency should lead to  the correct picture of drag and diffusion
of a heavy quark in the boundary theory. This idea has been sugested by 
several authors but was never clarified \cite{Rey:1998bq,Myers:2007we}.  
When analyzing dissipation in the context of black holes, the 
notion of a stretched horizon plays a central role \cite{Kovtun:2003wp,Parikh:1997ma,KSThorne}  and this surface 
is shown in \Fig{string1fig} at $r=r_h = 1 + \epsilon$.
 It will turn out that integrating out the fluctuations within 
the stretched horizon will yield a stochastic equation of motion
with the required noise at the horizon.

To keep this work self contained we have provided a
fairly extensive review of the necessary ingredients.
Section~\ref{Langevin} reviews the contour formalism 
of thermal field theory and shows how Brownian motion arises  
naturally from this algebraic structure. Section ~\ref{AdSContour} reviews
AdS/CFT in the Kruskal plane and generalizes the Kruskal/Keldysh correspondence to $\sigma=0$ and the $ra$ basis which is more natural for the classical 
AdS/CFT setup. Section ~\ref{Bulk_Stochastic_contour} shows that
bulk to bulk contour correlation functions respect KMS relations. 
This  result which, while known \cite{Gibbons:1976pt,Gibbons:1976es}, was not entirely appreciated. 
Finally, Section ~\ref{Bulk_brownian}
presents a derivation of the stochastic force on the horizon and
the subsequent section analyzes the results. 
A readable summary of the results 
and the bulk picture of Brownian motion is presented in Section ~\ref{Summary}.

\section{Langevin Dynamics}
\label{Langevin}

\subsection{Notation and Overdamped Motion }
\label{basic}

In this section we will review briefly the  
Langevin process. Consider a heavy particle moving with velocity $\dot x$,
subjected to drag $-\etav \dot x$ and random noise $\xi$
\st
\label{newton_langevin}
M_{\rm kin} \frac{d^2x}{dt^2} + \etav \frac{dx}{dt}  =  \xi  \nc
\qquad  \llangle \xi(t) \xi(t') \rrangle  =  2 T\etav \,\delta(t-t') \np
\stp
Here and below we will only write the equation of motion for
the $x$ component of the motion,  $\xi\equiv\xi_x$.
We also have anticipated that the strength of the noise 
$\llangle \xi(t) \xi(t')\rrangle$ is related to the drag
$\etav$ through the fluctuation 
dissipation theorem. $M_{\rm kin}$ is the quasi-particle mass  
including in-medium modification of the mass.
The diffusion coefficient  can be written as
\begin{equation}
\label{DfromKappa}
   D = \frac{T}{\etav}  \np
\end{equation}
Let us recall how this formula follows.  Consider the ultimate long time
limit. In this limit the $-M_{\rm kin}\omega^2$ term may be dropped
since it is proportional to frequency squared  and the 
motion is overdamped
\st
  \etav  \frac{dx}{dt} =   \xi  \nc
\stp
$i.e.$  the dissipation exactly balances the force. Solving for 
the position of the quark, 
 squaring, and averaging,  we find the squared displacement
\st
 \llangle x^2(t)\rrangle = 2 \frac{T}{\etav} t \np
\stp
In the diffusion equation   a Gaussian drop of dye spreads out
as $\llangle x^2(t) \rrangle = 2 D t$   leading to the identification 
in \Eq{DfromKappa}.  This overdamped limit where the 
drag force exactly balances the velocity will be central in discussing  
the membrane paradigm and black holes.  

\subsection{Langevin Dynamics From the Contour}
\label{contour_langevin}

\begin{figure}
\begin{center}
\includegraphics[width=4.0in]{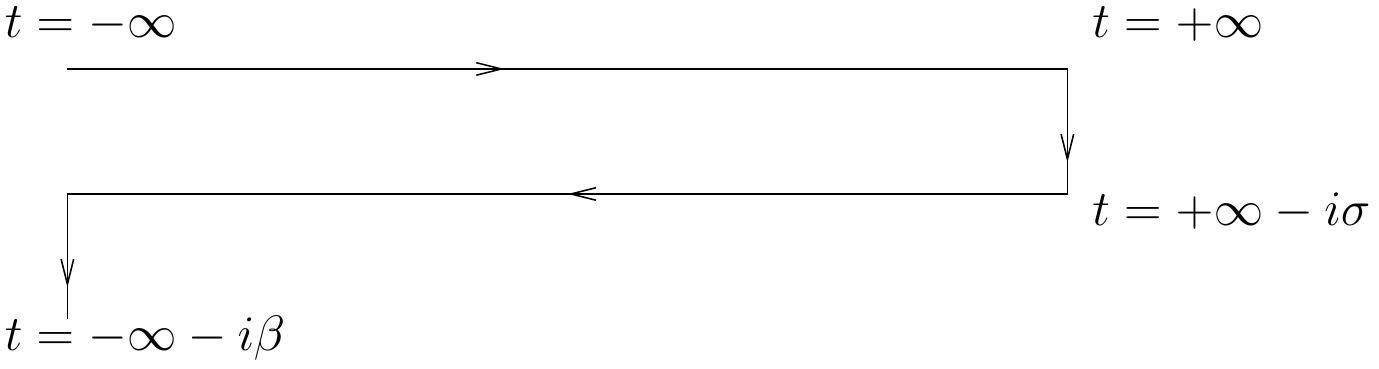}
\caption{The Schwinger Keldysh contour. The fields labeled by ``1" live
on the upper time ordered axis, while the fields labeled by ``2" live on 
the lower anti-time ordered axis. 
\label{Keldysh}
}
\end{center}
\end{figure}

Next we review how the Langevin equations can be derived
from the real time path integral \cite{Caldeira:1982iu,Feynman:1963fq}. 
The purpose here is not to track down every intricacy as there are reviews 
for this purpose \cite{Grabert:1988yt}.
In the recent literature we have found Refs.~\cite{Greiner:1998vd,Boyanovsky:2004dj}  instructive.
For a heavy particle 
coupled to an equilibrated  bath of forces  the real time partition function  is
\st
\label{ZHQ}
Z = \llangle \int [\Dt x_1] [\Dt x_2]\,  e^{i \int \dd t_1 M_Q^o \dot x_1^2  } \,
e^{- i\int \dd t_2 M_Q^o \dot x_2^2  }\,   e^{i\int dt_1 \F_1(t_1) x_1(t_1) - i\int dt_2 \F_2(t_2) x_2(t_2) } \rrangle_{\rm bath} \, ,
\stp
where the path integration  is along  the Schwinger-Keldysh contour
 \cite{Keldysh:1964ud,Schwinger:1960qe,Chou:1984es,Greiner:1998vd}. 
The path integral over the ``1" type coordinates  (the upper line in 
\Fig{Keldysh}) represents the amplitude,
while the path integral  over the ``2" type coordinates  (the lower line in \Fig{Keldysh}) represents the conjugate amplitude. The path integral over the vertical pieces of the contour (which
is not explicitly written in \Eq{ZHQ} but is implied by \Fig{Keldysh})
represents the average over the thermal density matrix $e^{-\beta H}$. 
The choice of $\sigma$ is arbitrary and reflects the fact that the 
thermal density matrix is stationary and therefore the average can be 
performed either at future or past infinity. An analogous dichotomy 
exists in the gravitational setup as discussed in \Sect{AdSContour}.
We have not written the path integral for the bath which is also evaluated along the 
contour.

When the particle is very heavy, the action is large and the motion is quasi-classical. The medium forces are small compared to the inertial terms and 
we can expand to second order, average over the bath,  
 and re-exponentiate to find 
\st
Z = \int  [\Dt x_1 ][\Dt x_2] 
e^{i\int \dd t_{1} M^{o}_Q \dot x_{1}^2  } \, e^{-i\int \dd t_2  M^{o}_Q  \dot x_2^2  }
e^{-\frac{1}{2}  \int \dd t\dd t' x_s(t) \left[\llangle \F(t) \F (t')\rrangle\right]_{ss'} x_{s'}(t')  }  \np
\stp

Here the  matrix of contour ordered correlation functions is
\st
  \left[\llangle \F(t) \F(t') \rrangle \right]_{ss'}  
\equiv i \left[ \begin{array}{cc}
\phantom{+} G_{11}(t,t') &  -G_{12}(t,t')  \\
-G_{21}(t,t') &  \phantom{+} G_{22}(t,t') \end{array} \right] \nc
\stp
where for instance $G_{12}(t,t')=\llangle F_1(t) F_2(t') \rrangle$ 
is the average of the forces over the partition function of the bath.
The relation to the operator formalism is the following: 
we define  the time dependent  operators
\st
\hatf_1(t) = e^{iH t} \hatf(0) e^{-iH t} \nc \qquad \hatf_2(t) = e^{+i H(t-i\sigma) } \hatf(0) e^{-iH(t-i\sigma) } \nc
\stp
and then the correlation functions are
\begin{subequations}
\bg
iG_{11}(t,t') &=& \llangle\, T\, \hatf_1(t)\, \hatf_1(t')\,\rrangle \, ,    \\
iG_{12}(t,t') &=& \llangle\, \hatf_2(t')\, \hatf_1(t) \,\rrangle \, , \\
iG_{21}(t,t') &=& \llangle\, \hatf_2(t)\, \hatf_1(t') \,\rrangle \, , \\
iG_{22}(t,t') &=& \llangle\, \tilde{T}\, \hatf_2(t)\, \hatf_2(t') \, \rrangle \, .
\nd
\end{subequations}

The KMS condition relates the different time orderings  
so there is really only one independent function 
which can be taken to be the retarded Green function 
\st
 iG_R(t)= \theta(t) \llangle[\hatf(t),\hatf(0)]\rrangle_{\rm bath}  \np
\stp
Using completeness and KMS relations  it can be shown that 
\begin{subequations}
\label{KMS_ugly}
\begin{align}
iG_{11}(\omega)
=& 
+i\Re\, G_R(\omega) - (1 + 2n)  \Im\, G_{R}(\omega) \nc  \\
iG_{22}(\omega) 
=&
-i\Re\, G_R(\omega) - (1 + 2n)  \Im\, G_{R}(\omega) \nc  \\
iG_{12}(\omega) 
=&
-2 n e^{\omega \sigma} \Im\, G_{R}(\omega) \nc \\
iG_{21}(\omega)
=&
-2(1+n) e^{-\omega \sigma} \Im\, G_{R}(\omega)  \np
\end{align}
\end{subequations}

Hidden in these relations is  the inter-relation between 
the drag and noise. 
Almost all applications of real time thermal field theory have relied upon a rewritten version of the 
path integral known as the $ra$ formalism (see, $e.g.$, Refs.~\cite{Chou:1984es,Greiner:1998vd,CaronHuot:2008uh, Mueller:2002gd}). 
(Of particular note is the next to leading order computation 
of the heavy quark diffusion coefficient in weakly coupled 
$\N=4$ SYM which relied exclusively on the $ra$ setup \cite{CaronHuot:2008uh}.)   Here and 
below we take  $\sigma=0$ 
 and the average over 
the initial density matrix is entirely in the past.  
Since in a quasi 
classical limit the amplitude is not very different from the conjugate 
amplitude we define the retarded ($r$) and advanced fields ($a$) for the particle and also for the forces
\st
  x_{r} = \frac{x_1 + x_2}{2} \nc  \qquad x_a = x_1- x_2 \nc
  \qquad \qquad \F_{r} = \frac{\F_1 + \F_2}{2} \nc \qquad \F_a = \F_1 - \F_2 \np
\stp
$x_a$ should be considered a small parameter in the classical limit \cite{Mueller:2002gd}. 
Effecting  this transformation we find
\st
Z = \int [\Dt x_r] [\Dt x_a] e^{-i \int \dd t\, 
 M_Q^o x_a \ddot x_r   -  
\int \dd t \dd t'  x_a(t) iG_{R}(t,t') x_r(t') 
- \frac{1}{2} x_{a}(t) G_{\rm sym}(t,t') x_a(t') } \np
\stp
We have  defined the  propagators
\begin{subequations}
\bg
  G_{\rm sym}(t,t') = 
\llangle \F_r(t) \F_r(t') \rrangle &=& 
\frac{1}{2} \llangle \{\hatf(t),\hatf(t')\} \rrangle \nc \\
  iG_{R}(t,t') = \llangle \F_r(t) \F_a(t') \rrangle &=& 
\theta(t) \llangle [ \hatf(t), \hatf(0) ] \rrangle \nc \\
  iG_{A}(t,t') = \llangle \F_a(t) \F_r(t') \rrangle  &=& 
-\theta(-t) \llangle [ \hatf(t), \hatf(0) ] \rrangle  \nc
\nd
\end{subequations}
made use of the familiar  contour relationships
\begin{subequations}
\bg
  G_{\rm sym} &=& \frac{i}{4} \left[G_{11}  + G_{22} + G_{12} + G_{21} \right] \nc  \\ 
  iG_{R} &=& \frac{i}{2} \left[ G_{11} - G_{22} + G_{21} - G_{12} \right] \nc \\
  iG_{A} &=& \frac{i}{2} \left[ G_{11} - G_{22} - G_{21} + G_{12} \right] \nc \\
     0  &=&  G_{11} + G_{22} - G_{12} - G_{21}  \nc
\nd
\end{subequations}
and also have used the  
reality relation between the advanced and retarded propagators
\st
  G_{A}(t)  = G_{R}(-t) \nc   \qquad   G_{A}(\omega) = G_{R}^*(\omega)  \np
\stp
The KMS relations become simply the canonical form of the 
fluctuation dissipation theorem
\st
     G_{\rm sym}(\omega) = -(1 + 2n) \Im G_{R}(\omega)  \np
\stp
In Fourier space the path integral is simply
\st
\label{rapathintegral}
Z = \int  [\Dt x_r ][\Dt x_a] 
\exp\left( -i\int \frac{\dd \omega}{2\pi} x_{a}(-\omega)[-M^{o}_Q\omega^2 + G_{R}(\omega)] x_r(\omega)  \right) \, 
e^{-\frac{1}{2} \int \frac{\dd \omega}{2\pi} x_a(-\omega)[ 
G_{\rm sym}(\omega) ] x_{a}(\omega)  } \np
\stp
After Fourier transforming the  Gaussian by introducing a noise variable, 
\st
e^{-\frac{1}{2} \int \frac{\dd \omega}{2\pi} x_a(-\omega)[ 
G_{\rm sym}(\omega) ] x_{a}(\omega)  } = \int [\Dt \xi] e^{+i\int x_a(-\omega) \xi(\omega)}\, e^{-\frac{1}{2} \frac{\xi(-\omega) \xi(\omega)}{G_{\rm sym}(\omega)} } \nc
\stp
the partition function reads
\bg
\label{rapartition_function}
Z &=& 
\int  [\Dt x_r ][\Dt x_a][\Dt \xi] 
e^{-\frac{1}{2} \int \frac{\dd \omega}{2\pi} \frac{\xi(-\omega) \xi(\omega) }{G_{\rm sym}(\omega) } }\,  \nonumber \\
  & & \qquad \times 
\exp\left(-i\int \frac{\dd \omega}{2\pi} x_{a}(-\omega)\left [-M^{o}_Q\omega^2 x_{r}(\omega)  + G_{R}(\omega) x_r(\omega) -  \xi(\omega) \right] \right)    \np
\nd

At this point one may integrate over $x_a(-\omega)$ yielding the path 
integral
\st
Z = \int  [\Dt x_r ][\Dt \xi] 
e^{-\frac{1}{2} \int \frac{\dd \omega}{2\pi} \frac{\xi(-\omega) \xi(\omega) }{G_{\rm sym}(\omega) } }   
\delta_{\omega} \left[ -M^{o}_Q \omega^2 x_r(\omega) + G_{R}(\omega) x_r(\omega)  - \xi(\omega) \right] \np
\stp
This equation means that the partition function is simply an 
average over the classical trajectories under the influence 
of a  random  colored force,
\st
\label{GeneralizedLangevin}
\left[-M_{Q}^{o}\omega^2 + G_{R}(\omega) \right] x(\omega) = \xi(\omega) \nc \qquad   \llangle \xi(-\omega) \xi(\omega) \rrangle 
=  G_{\rm sym}(\omega) = -(1 + 2n) \,\Im G_R(\omega)  \np
\stp
In time this equation reads
\st 
\label{GeneralizedLangevinT}
M_{Q}^{o}\frac{d^2x}{dt^2}  + \int^t \dd t'\, G_{R}(t,t') \,  x(t') = \xi(t)\nc \qquad   \llangle \xi(t) \xi(t') \rrangle =  G_{\rm sym}(t,t') \nc
\stp
which is a generalized Langevin equation with the drag and corresponding noise
\cite{Forster, BooneYip}.
One method to implement such colored noise on the 
computer has been given in \Ref{Xu:1999aq}.

At small frequencies we can  expand the retarded Green function
to $\omega^2$ 
\st
   G_{R}(\omega) = - i\omega \etav  - \Delta M \omega^2 \nc
\stp
and then the effective equation of  motion is the original Langevin equation
\st
M_{\rm kin} \frac{d^2x(t)}{dt^2}  + \etav \frac{dx(t)}{dt} = \xi(t) \nc \qquad \llangle \xi(t) \xi(t') \rrangle =  2 T\etav\, \delta(t-t')  \nc
\stp
where we have defined the kinetic mass as $M_{\rm kin}(T) = M_Q^o + \Delta M $.

\section{Review of Trailing Strings}
\label{TrailingStrings}

Our purpose here is to collect some of the results from the 
AdS/CFT correspondence on the drag and diffusion of heavy quarks \cite{HKKKY,Jorge,Gubser}.
A canonical choice of coordinates  for the  metric of 
black hole ${\rm AdS}_5$ (which we will  denote with bars)  is
\st
\label{canonical}
ds^2_5 =  \frac{\bar{r}^2}{L^2} \left[ - \f(b \bar r) dt^2 +  d\x^2 \right] +  \frac{ L^2 d\bar r^2}{\f(b\bar r) \bar r^2}  \nc
\stp
with  $f(r) = 1-1/r^4$.  Here $b$ is the inverse  horizon radius  which is related 
to the Hawking temperature, $b = 1/\pi T L^2$. 
We will use a different set of conventions defining   $r\equiv b\bar{r}$,
such that $r$ is a measure of energy in units of $\pi T$
\st
ds^2 =  (\pi T)^2 L^2 \left[ -r^2 \f(r) dt^2 +  r^2 d\x^2 \right] +  \frac{ L^2 dr^2}{\f(r) r^2}  \np
\stp

We are considering the dynamics of very massive quark which is 
represented in AdS$_5$ by a long string stretching from 
the horizon upwards towards the  AdS boundary terminating
at $r_{m}$ as illustrated in \Fig{string1fig}.  This straight
string is a solution to the classical Nambu-Goto equations of
motion. Then we consider small fluctuations around this long 
straight string.

The action for these fluctuations is (see \app{fluctuations})
\st
\label{action_body}
S = - \int \dd t \dd r \left[\m + \frac{1}{2}\T(r) (\partial_rx)^2 
 - \frac{\m}{2\f}(\partial_t x)^2 \right] \nc
\stp  
where  
\st
\label{Tofr}
  \T(r) = \frac{(\pi T)^3 L^2}{2\pi\ls2}\,  fr^4   =  (\sqrt{\lambda} \pi^2 T^3/2) \,  fr^4 \nc
\stp 
has the meaning of the local tension,
and  
\st
   \m = \frac{(\pi T) L^2}{2\pi \ls2} = \frac{\sqrt{\lambda} T}{2} \nc
\stp
is mass per unit $r$.
The zero temperature  mass of the quark  $M_{Q}^o=\m r_m$. The 
speed of waves on the string is $c_s = \sqrt{T_o(r)f/m} = \pi T fr^2$ 
and therefore waves  propagate from $r_h=1 + \epsilon$ to the boundary in a time of order $\sim 1/\pi T \log(1/\epsilon)$.

For a prescribed boundary value $x_{o}(\omega)$ we can
solve for the classical waves on the string by imposing retarded boundary 
conditions at the horizon. 
The classical equations of  motion after Fourier transforming in time are
\st
\label{eom_bulk}
 \partial_r\big( \T(r) \partial_r x(\omega,r) \big) + \frac{\m \omega^2}{f}x(\omega,r)  = 0 \np
\stp
We will define $F_{\omega}(r)$ as the retarded boundary to bulk 
propagator, $i.e.$ the solution which 
satisfies 
\[
\lim_{r\rightarrow r_m} F_{\omega}(r)=1 \nc
\]
and retarded boundary conditions.

The waves on the string due to the sinusoidal motion of the 
end point of the string are 
\st
\label{trailing_string}
  x_o(\omega,r) = x_o(\omega) F_{\omega}(r) = x_o(\omega) + \frac{-i\omega x_{o}(\omega)}{(2\pi T)} \left[\tan^{-1}\left(z\right) - \tanh^{-1}\left(z\right) \right] + O(\omega^2)  \nc
\stp
with $z\equiv1/r$.
The term  multiplying the velocity $-i\omega x_o(\omega)$ 
is the ``trailing string" solution of Refs.~\cite{Gubser,HKKKY}. (In writing this 
equation we have assumed that we are not exponentially close to the horizon, $i.e.$ $\omega \log(1/\epsilon) \ll \pi T$.) Thus to lowest 
order the string trails behind the sinusoidal motion according to the expected 
form. The next term in this series is proportional to the 
acceleration and is analyzed in \app{fluctuations}.

The retarded force-force correlator
is found by taking the boundary limit \cite{Policastro:2002se,Jorge} 
\st
G_{R}^{o} \equiv \lim_{r\rightarrow r_m} \T(r) F_{-\omega} \partial_r F_{\omega} = -M_{Q}^{o} \omega^2 + G_{R}(\omega)  \nc
\stp
where $M_{Q}^{o}$ is the zero temperature quark mass, and 
the  term $-M_{Q}^{o}\omega^2$  arises from the ``divergent" part 
of the boundary limit.  Then using $\T(r)$ given above and the retarded solution given in 
\Eq{trailing_string} and \app{fluctuations}, we find 
\st
G_{R}(\omega) = -i\omega\etav - \Delta M \omega^2 \nc
\stp
with 
\bg
   \etav  &=& \frac{1}{2} \sqrt{\lambda} \pi T^2 \nc \\
   \Delta M &=& -\frac{\sqrt{\lambda} T}{2} \np
\nd
As discussed above, the field theory interpretation
from this form for the retarded force-force correlator is
that the quark obeys the stochastic motion in \Eq{GeneralizedLangevinT} with
the specified transport coefficient $\etav$ (first computed in \Ref{HKKKY,Jorge,Gubser}) and in medium 
mass shift $\Delta M$ (first computed in \Ref{HKKKY}). The primary 
aim of this work is to show how this stochastic equation is 
derived in AdS/CFT 
and to give a bulk picture to the stochastic process. 

Before taking up this enterprise we make the following technical notes 
about the solutions to the bulk equations of motion \Eq{eom_bulk}:
\begin{enumerate}
\item The conjugate solution satisfies 
$F^{*}_{\omega} = F_{-\omega}$ and obeys advanced boundary 
conditions. When constructing solutions we note that
$\Im F_{\omega}(r)$ obeys the same differential equation 
but is normalizable, $i.e.$ $\Im F(r) \rightarrow 0$ as $r\rightarrow \infty$.
\item The imaginary part of the retarded  Green function, 
\st
\label{ImGrindep}
 \Im G_{R}(\omega) =  \frac{\T(r)}{2i} \left[ F_{-\omega}(r) \partial_r F_{\omega}(r)   - F_{\omega}(r) \partial_r F_{-\omega}(r) \right]  \nc
\stp
can be evaluated at any radius. This is because the term in 
square brackets is the Wronskian of the differential equation 
which depends on $r$ in a particular way 
which precisely cancels the leading $\T(r)$ factor. This fact will be used repeatedly in \Sect{analysis}.
\item When evaluating the evolution of perturbations in the bulk, it is
often very useful  to define the retarded bulk to bulk propagator.
Many authors  ({\it e.g.} Refs.~\cite{Chesler:2007sv,Chesler:2008wd})
have defined a Green function  
which is infalling at the horizon and normalizable at the boundary 
\st
\label{gret}
  G_{\rm ret}(\omega,r,\bar r) = \frac{\Im F_{\omega}(r) F(\rbar) \theta(r,\rbar) + F_\omega (r) \Im F_{\omega}(\rbar)\theta(\rbar,r) }{\T(\rbar) W_{\rm ret}(\rbar)} \nc
\stp
where we have defined the retarded Wronskian$\times \T(r)$
\begin{align}
\label{retw}
 \T(r) W_{\rm ret}(r) =& \T(r)  \left[ \Im  F'(r) F(\bar r) - F'(r) \Im F(r) \right] \nc \nonumber \\
             =& -\Im G_{R}(\omega) \np
\end{align}
Note again this combination
is proportional to the retarded force-force correlator 
and is independent of $r$.
\end{enumerate}

\section{The Boundary Picture of Stochastic Motion}
\label{Boundary Stochastic}

\subsection{AdS/CFT on the Contour}
\label{AdSContour}

In this section we review how the Schwinger-Keldysh formalism 
is constructed in AdS/CFT by using the full Kruskal structure
of the black hole. Here 
we will extend the
results of Ref.~\cite{Herzog:2002pc}  only slightly 
to show how the arbitrary  $\sigma$ of the contour
is present in the Kruskal formalism.
We will then 
specialize to $\sigma=0$ and indicate how the results 
appear in the $ra$ basis.
The Kruskal/Keldysh correspondence was based on the pioneering work on Hawking radiation 
\cite{Hawking:1974sw,Hartle:1976tp,Unruh:1976db,Israel:1976ur,Gibbons:1976pt,Gibbons:1976es}. 

\begin{figure}
\begin{center}
\includegraphics[height=2.5in]{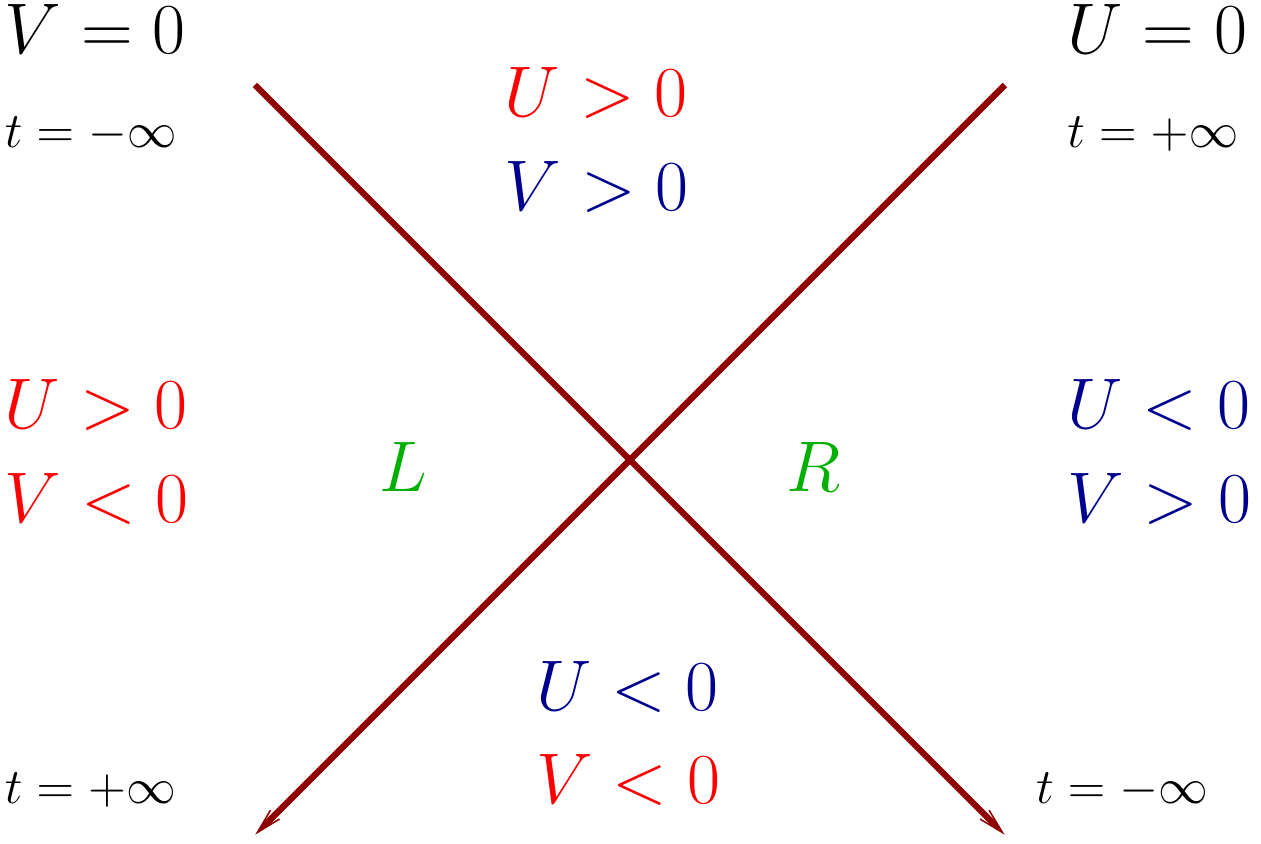}
\caption{The full Kruskal plane. The right quadrant corresponds
to the amplitude of the field theory (the ``1" axis) while the left quadrant 
corresponds to the conjugate amplitude of the field theory (the ``2" axis). }
\label{KruskalFig}
\end{center}
\end{figure}

The Kruskal plane 
of the eternal black hole is 
exhibited in \Fig{KruskalFig} (see \app{Krusk_notation} for a summary of
the Kruskal conventions adopted here.) Outside of the horizon there are  two
causally disconnected space time geometries both of which are
asymptotically AdS. The right quadrant corresponds to 
the amplitude of the CFT while the left quadrant corresponds
to the conjugate amplitude of the CFT.  The dynamics
of the two AdS/CFT's are coupled through boundary conditions
at past and future infinity. 

AdS/CFT in practice amounts to a four step procedure. 
\begin{enumerate}
\item First determine a solution to the classical equations
of motion  with certain boundary values.
\item 
In order to uniquely specify this solution one must specify 
horizon boundary conditions.
\item This solution is then substituted into the  
classical action which reduces to a boundary term.
\item 
Finally, variation with respect  to the boundary 
values determines the correlation functions of the CFT. 
\end{enumerate}
In the full Kruskal plane 
these standard AdS/CFT steps are the following. \\ 
\newline
{\noindent 1.} First  the equations of 
motion  for the fluctuations on the string 
are solved in the full Kruskal plane with the boundary values
\begin{subequations}
\label{boundary_constraints}
\bg
 \lim_{r_1\rightarrow r_{m}} x(\omega,r_1) &=& x_1^{o}(\omega) \nc \\
 \lim_{r_2 \rightarrow r_{m}} x(\omega,r_2) &=& x_2^{o}(\omega) \np
\nd
\end{subequations}
Here  $r_1$ and $r_2$ are the radial coordinates in 
the right and left quadrants respectively.
The general solutions in the right and left quadrants are 
\bg
\label{xofr1}
 x(\omega,r_1) &=& a(\omega) F_{\omega}(r_1) + b(\omega) F_{\omega}^{*}(r_1) \nc \\
 x(\omega,r_2) &=& c(\omega) F_{\omega}(r_2) + d(\omega) F_{\omega}^{*}(r_2) \np
\nd
Since we have only two boundary constraints given by \Eq{boundary_constraints}, 
to uniquely specify the solution we must specify horizon boundary conditions.

{\noindent 2.} The horizon boundary conditions specify the connection between the
solution in the right quadrant and the solution in the left quadrant. 
We note that near the horizon the retarded and advanced solutions behave as
\bg
  e^{-i\omega t} F_{\omega}(r_1) &\sim& e^{-i\frac{\omega}{2\pi T} \log(V) } 
  \nc \\
  e^{-i\omega t} F_{\omega}^{*}(r_1) &\sim& e^{+i\frac{\omega}{2\pi T} \log(-U) }   \np 
\nd
The horizon conditions proposed in \Ref{Herzog:2002pc} are that the 
solution should be analytic in the lower $V$ plane and upper $U$ plane. This 
physical choice is based on the intuition that if the right 
universe is to represent the time-ordered amplitude of the CFT,
then the infalling mode ($F_{\omega}$) 
should be positive energy (analytic in in lower $V$) while the outgoing
mode ($F^{*}_{\omega}$) should  be  negative energy (analytic in upper $U$).
For instance in a free scalar theory the Green function at zero temperature 
is
\st
iG_{11}(t,t') = \frac{1}{2E_\p} e^{-iE_\p(t-t')}\theta(t,t') + \frac{1}{2E_\p} e^{+iE_\p (t-t')} \theta(t',t)  \nc
\stp
and therefore a one type source $J_1(t')$ will produce fields with 
positive energy wave at later time and induce negative energy wave 
at an earlier time.
With this choice Herzog and Son analytically continue from the right quadrant ($V>0$, $U<0$)  
to the left quadrant ($V<0$, $U>0$) yielding from the solution 
in the right quadrant $x(\omega,r_1)$ the
solution in the left quadrant 
\st
\label{xofr2}
  x(\omega, r_2) = a(\omega) \, e^{-\omega/2T }\, F_{\omega}(r_2)  + b(\omega) 
            \, e^{+\omega/2T} F^{*}_{\omega}(r_2)  \np
\stp
Now the values of $a(\omega)$ and $b(\omega)$ can be determined from 
the boundary values $x_1^{o}(\omega)$ and $x_2^{o}(\omega)$. 
The resulting solution ultimately reproduces the  contour
correlation functions for the specific choice of $\sigma = \beta/2$ 
\cite{Herzog:2002pc}. 

Here we will generalize their work slightly by extending 
$V \rightarrow |V|e^{-i\theta}$ and $-U \rightarrow |U|e^{-i(2\pi -\theta)}$;
 in the original work $\theta=\pi$.
With this choice for the analytic continuation of $U$ and $V$, the radius
is fixed since $(-U)V$ is changed ultimately by a factor $e^{-2\pi i}$ 
(note \Eq{UVrelation}). We 
will find that this choice for the analytic continuation reproduces
the contour relations for arbitrary $\sigma$. Loosely speaking, the 
difference between these choices is a decision about whether to 
perform the thermal average at past infinity $V=0$ or future infinity $U=0$.
We do not have a sharper physical explanation for this choice at this point.
With this analytic continuation the infalling and outgoing  solutions 
behave as
\bg
\label{analytic_continuation}
   F_{\omega}(r_1)   & \rightarrow & e^{-\omega \sigma} F_{\omega}(r_2) \nc \\
   F^{*}_{\omega}(r_1) & \rightarrow & e^{+\omega/T} e^{-\omega \sigma} F^{*}_{\omega}(r_2) \nc
\nd
where we have defined\footnote{Of course this definition will 
later turn out to correspond to the usual field theory definition.}
\st
   \sigma = \frac{\theta}{2\pi T} \np
\stp
The solution in \Eq{xofr1} when analytically continued to the left quadrant 
behaves as
\st
\label{xofr2b}
  x(\omega, r_2) = 
a(\omega) \, e^{-\omega \sigma} F_{\omega}(r_2)  + 
  b(\omega) \, e^{+\omega/T} e^{-\omega \sigma} F_{\omega}^{*}(r_2)  \np
\stp
Now as before we can solve for the coefficients $a(\omega)$ and
$b(\omega)$ in terms of the  $x_1^{o}(\omega)$ and $x_2^{o}(\omega)$
yielding the   result 
\begin{subequations}
\bg
a(\omega) &=&   x_1^o({\omega}) (1 + n(\omega)) - x_2^o(\omega) e^{\omega \sigma} n(\omega)  \nc \\
b(\omega) &=& 
 x_{2}^o({\omega}) e^{\omega \sigma}n(\omega)  - x_1^{o}(\omega) n(\omega) \np
\nd
\end{subequations}
The solution is now fully specified by its right and left quadrant solutions, \Eq{xofr1} and \Eq{xofr2b}. \\
\newline
{\noindent 3.} 
Now we can substitute the solution into the boundary action 
\st
\label{bndry12action}
S_{\rm bndry} = - \frac{\T(r_m)}{2} \int_{r_1} \frac{\dd\omega}{2\pi}  
x_1(-\omega,r_1) \partial_r x_1(\omega,r_1) + 
 \frac{\T(r_m)}{2} \int_{r_2} 
\frac{\dd\omega}{2\pi} x_2(-\omega,r_2) \partial_r x_2(\omega, r_2)  \nc
\stp
and determine  the generating function
\begin{align}
\label{is12bndry}
iS_{\rm bndry} = -\frac{1}{2} \int \frac{\dd\omega}{2\pi} & x_{1}^{o}(-\omega) \left[ +i\Re G_R^{o} - (1 + 2n) \Im G_R^{o} \right] x_{1}^{o}(\omega) \nonumber \\
   + \,& x_{2}^{o}(-\omega) \left[ -i\Re G_R^{o} - (1 + 2n) \Im G_R^{o} \right] x_2^{o}(\omega) \nonumber\\
   - \, & x_1^{o}(-\omega) \left[-2n  e^{+\omega\sigma} \Im G_{R}^{o} \right] 
x_{2}^{o}(\omega) 
  - x_2^{o}(-\omega) \left[ -2(1 + n) e^{-\omega\sigma} \Im G_R^o \right] x_{1}^{o}(\omega) \np
\end{align}
Here we identified  the retarded Green function with
\st
 G_{R}^{o} (\omega) =
\left.  \T(r) \frac{F_{-\omega}(r) \partial_r F_{\omega} }{|F(\omega,r)|^2}  
\right|_{r=r_m} \np
\stp

{\noindent 4.} Now by taking variations with respect to 
 $x_{1}^{o}$ and $x_{2}^{o}$ one can reproduce the full 
contour correlation functions which obey the appropriate KMS relations in 
\Eq{KMS_ugly}.

\subsection{Simple Results in the $ra$ Setup}

Before closing this review of Ref.~\cite{Herzog:2002pc}  let us show
how these results work out in the $ra$ basis. Specializing 
to $\sigma=0$ and introducing the bulk $r$ and $a$ fields
\st
 x_{r}(\omega,r) = \frac{x_1(\omega,r) + x_2(\omega,r) }{2}\nc \qquad
 x_{a}(\omega,r)= x_{1}(\omega,r) - x_2(\omega,r)  \np
\stp
we find that the bulk fields are rather simply related to the 
boundary fields  in the $ra$ setup
\begin{subequations}
\label{sol_ra}
\begin{align}
  x_{a}(\omega,r) =& F^{*}_{\omega}(r)  \, x^{o}_{a}(\omega) \nc\\
  x_{r}(\omega,r) =& F_{\omega}(r)\, x^{o}_r(\omega) + i(1 + 2n) \,\Im F_{\omega}(r) x_{a}^{o}(\omega) \np
\end{align}
\end{subequations}
The boundary action in the $ra$ formalism after rewriting \Eq{bndry12action}  is
\st
S_{\rm bnry} = - \frac{\T(r_m)}{2}  \int_{r_m}  \frac{\dd\omega}{2\pi}\, 
 \left[ x_{a}(-\omega,r) \partial_r x_r(\omega,r) +  x_{r}(-\omega,r) \partial_r x_a(\omega,r)  \right] \np
\stp 
Substituting the solutions given in \Eq{sol_ra} gives the desired result in 
the  $ra$ setup
\st
iS_{\rm bnry} = -i\int\frac{\dd \omega}{2\pi} x_a^{o}(-\omega) \left[ G_{R}(\omega) \right] x_{r}^{o}(\omega)  - \frac{1}{2} \int \frac{\dd\omega}{2\pi} x^{o}_{a}(-\omega)\left[G_{\rm sym}(\omega)\right] x^{o}_{a}(\omega) \nc
\label{Sbnry}
\stp
with  $G_{\rm sym}(\omega) = -(1 + 2n) \Im G_R(\omega) $.
Of course this could be found simply by changing variables  in 
\Eq{is12bndry}, but  it is hoped that the relatively simple causal relations
in \Eq{sol_ra} will be useful in understanding non-thermal thermal 
fluctuations of  AdS/CFT in an 
out of equilibrium setting \cite{Chesler:2008hg}.

\subsection{Stochastic Motion on the Boundary }

Now we will show how these results lead directly to stochastic motion 
on the boundary.  A formal path integral for the string is
\st
 Z = \int \left[\Dt x_1^{o}\, \Db x_1\right] 
          \left[\Dt x_2^{o} \,\Db x_2\right] e^{iS_1 - iS_2} \nc 
\stp
where in the right quadrant of the 
Kruskal plane, $\Dt x_1^{o}$ is a temporal path for
the string endpoint,
$\Db x_1$ is a bulk path integral for the body of the string,
and the analogous ``2" symbols are defined for the left quadrant.
Slightly more explicitly
\st
 \left[\Dt x_1^o\right] =  \prod_{t} \dd x_1^{o}(t) \nc \qquad  
[\Db x_1] = \prod_{t,r} \dd x(t,r)  \np
\stp
Imagine integrating out the bulk coordinates to find an effective 
action for the string endpoints. In a Gaussian approximation the 
integrals can be done. The result is a determinant times the 
exponential of the action evaluated 
with the classical solution which passes through the two endpoints  
and which obeys the appropriate horizon boundary conditions.  
This is of course simply the boundary action
\st
S_{\rm eff}^{o} = - \frac{\T(r_m)}{2} \int_{r_1} \frac{\dd\omega}{2\pi}  
x_1(\omega,r_1) \partial_r x_1(\omega,r_1) + 
 \frac{\T(r_m)}{2} \int_{r_2} 
\frac{\dd\omega}{2\pi} x_2(-\omega,r_2) \partial_r x_2(\omega, r_2)  \nc
\stp
which can be evaluated using the results of the previous section 
\st
iS_{\rm eff}^{o} = -i \int \frac{\dd \omega}{2\pi} x_a^{o}(-\omega) \left[ - M_{Q}^o\omega^2 +  G_{R}(\omega) \right] x_{r}^{o}(\omega)  - \frac{1}{2}\int \frac{\dd\omega}{2\pi} \,  x^{o}_{a}(-\omega)\left[ G_{\rm sym}(\omega) \right] x^{o}_{a}(\omega) \np
\stp
Now the partition function for the string  endpoint  is
\st
Z = \int \left[\Dt x_r^{o} \Dt x_a^o \right] e^{iS_{\rm eff}^{o} } \np
\stp
At this point we recognize the same partition function as discussed in the introduction, and following  
the discussion  after \Eq{rapartition_function} we conclude that
the string endpoint obeys the expected equations of motion
\st
\left[-M_{Q}^{o}\omega^2 + G_{R}(\omega) \right] x^{o}(\omega) = \xi^o(\omega) \nc \qquad   \llangle \xi^o(-\omega) \xi^o(\omega) \rrangle 
=  G_{\rm sym}(\omega) = -(1 + 2n) \,\Im G_R(\omega)  \np
\stp

\section{The Bulk Picture of Stochastic Motion}
\label{Bulk_Stochastic}

In the previous Section we integrated out all degrees of freedom of
the string, except for the endpoints, to obtain the effective action
for the endpoint coordinates.  In this Section we will integrate out
only the degrees of freedom inside a stretched horizon.  In this way
we will obtain an equation of motion for a string with a friction and
a noise acting on the string at the stretched horizon.

\subsection{Bulk to Bulk Correlators in Contour AdS/CFT}
\label{Bulk_Stochastic_contour}

In this section we wish to compute the bulk to bulk correlators 
in the Kruskal formalism of AdS/CFT. For simplicity consider correlation
functions of disturbances along an infinitely long string with
fixed endpoints stretching from the horizon to the boundary.
The action for these fluctuations has been given in \Eq{action_body}. 
We will consider the generating function
\st
Z[J_1,J_2] = \int [\Db x_1] [\Db x_2] \; e^{iS_1 - iS_2} \; e^{i \int {\dd t_1}{\dd r_1}  J_1(t_1,r_1) \, x_1(t_1,r_1)} \, e^{-i\int {\dd t_2}{\dd r_2} \, J_2(t_2,r_2) \, x_2(t_2,x_2) } \nc
\stp
where as before $[\Db x_1]$ indicates a bulk path integral.
We wish to compute all bulk two point functions
\st
 \left[iG(\omega,r,\rbar) \right]_{ss'} \equiv \int \dd t \, e^{i\omega t} \,
 \left[iG(t,r,\bar r) \right]_{ss'} \nc
\stp
with
\st
  \left[iG(t-\bar t,r,\bar r)\right]_{ss'} \equiv  \frac{1}{i^2} \frac{\delta^2 \ln Z }{\delta J_s(t,r) \delta J_{s'}(\bar t,\bar r) } 
\equiv i \left[ \begin{array}{cc}
G_{11} &  -G_{12}  \\
-G_{21} &  G_{22} \end{array} \right] \np
\stp

To this end we note that the classical equations of motion are
\st
 \left[\partial_r \big(\T(r) \partial_r x(\omega,r) \big)  + 
\frac{m\omega^2}{f} x(\omega,r) + J(\omega) \right] = 0 \nc
\stp
where $J(\omega)$ is the external force per unit length on the string.
Our task is to construct a Green function   of this differential equation
\st
 \left[\partial_r \big(\T(r) \partial_r G(r,\rbar) \big)  + 
\frac{m\omega^2}{f} G(r,\rbar) \right] = \delta(r-\rbar)  \np
\stp
Given two solutions of the differential equation 
$g_{>}(r)$ and $g_{<}(r)$  the Green function is easily constructed 
\st
\label{generalg}
  G(r,\bar{r}) = \frac{g_{>}(r) g_{<}(\bar r) \theta(r,\bar r) + g_<(r) g_>(\bar r) \theta(\bar r, r)  }{\T(\bar r) W(\bar r) } \nc 
\stp
where  the Wronskian is
\st
\label{wronskg}
 W(\bar r) = g_{>}'(\bar r)g_{<}(\bar r) - g_{<}'(\bar r) g_{>}(\bar r) \np
\stp

In the present case we wish to construct the Green function
in the full Kruskal plane.
We should find two solutions which obey the boundary conditions.  
These boundary conditions are 
that $g_{>}(r)$ should be normalizable as $r \rightarrow  \infty$ in the 
right quadrant of the Kruskal plane. Similarly
$g_{<}(r)$ should be normalizable as $r \rightarrow \infty$ in the left 
quadrant of the Kruskal plane.  
In formulas
\st
\label{gg1}
g_{>}(r_1) = \Im F_\omega(r_1)    \qquad \mbox{(right quadrant) } \nc
\stp
and 
\st
g_{<}(r_2) = \Im F_\omega(r_2) \qquad \mbox{(left quadrant) } \np
\stp
Using the boundary conditions of the real time AdS/CFT 
we extend across the horizon in the Kruskal plane. 
Writing $\Im F(r_1) = (F(r_1) - F^*(r_1))/2i$ and analytically continuing
$g_{>}$ according to \Eq{analytic_continuation} we have
\st
g_{>}(r_2) = \frac{1}{2i} 
\left[ e^{-\sigma \omega} F_{\omega}(r_2) - 
e^{\omega/T} e^{-\sigma \omega } F_{\omega}^*(r_2) \right] \qquad \mbox{(left quadrant)} \np
\stp
Similarly we realize that the analytic continuation of 
$g_<(r_2)$  to $r_1$ is
\st
\label{gl1}
g_{<}(r_1) = \frac{1}{2i} \left[ e^{+\sigma \omega}  F_{\omega}(r_1) - e^{-\omega/T} e^{\sigma\omega} F_\omega^{*}(r_1) \right] \qquad \mbox{(right quadrant)} \np
\stp

Now different correlators can be evaluated in a straightforward 
manner.  
Evaluating $G_{11}(r_1,\bar r_1)$ using \Eq{generalg}, \Eq{gg1}, and \Eq{gl1},
and for simplicity 
taking $r_1 > \bar{r}_1$,  we have
\begin{align}
 G_{11}(r_1,\bar{r}_1)=&  \frac{g_>(r_1) g_{<}(\bar{r_1}) }{W(\bar r_1) } \nc \nonumber \\
                      =& \Re G_{\rm ret}(r_1,\rbar_1) + i(1 + 2n)\Im G_{\rm ret}(r_1,\rbar_1)  \np
\end{align} 
Here $n(\omega) =1/(e^{\omega/T}  - 1)$ is
the thermal occupancy factor  and $G_{\rm ret}(r,\rbar)$ is
given in \Eq{gret}. 
In deriving this result we note the intermediate results
\begin{subequations}
\begin{align}
  W(\rbar_1) =&  e^{+\omega\sigma} (1 - e^{-\omega/T} ) \frac{1}{2i} W_\ret(\rbar_1) \nc  \\
  W(\rbar_2) =&  e^{-\omega\sigma} (e^{+\omega/T}-1) \frac{1}{2i} W_{\rm ret} (\rbar_2) \nc
\end{align}
\end{subequations}
and the much used identity $n(-\omega) = -(1 + n(\omega))$.

Continuing in this way, we  summarize these results for the bulk to bulk 
correlators 
\begin{subequations}
\begin{align}
iG_{11}(r,\bar r) = & +i\Re\, G_\ret(r,\rbar) - (1 + 2n)  \Im\, G_{\ret}(r,\rbar) \nc \\
iG_{22}(r,\bar r) = & -i\Re\, G_\ret(r,\rbar) -  (1 + 2n) \Im\, G_\ret(r,\rbar)  \nc \\
iG_{12}(r,\bar r) = & -2n  \, e^{+\omega \sigma} \Im\, G_\ret(r,\rbar) \nc \\
iG_{21}(r,\bar r) = & -2(1+n) \,   e^{-\omega \sigma} \Im\, G_\ret(r,\rbar)   \np
\end{align}
\end{subequations}
These are the familiar spectral and KMS relations between the retarded 
correlator and various contour correlation functions. It is reassuring
that these thermal relations between the bulk to bulk correlators arise
so easily in the Kruskal formalism although the result is not 
particularly new or surprising \cite{Gibbons:1976pt,Gibbons:1976es}. 

To show that these correlation functions are indeed correctly 
normalized, one can shift the fields in the 
usual way
\st
  x_{s}(\omega,r) = \delta x_s(\omega,r) -  \int\dd\rbar  \, \left[iG(\omega, r,\bar r) \right]_{ss'} J_{s'}(\omega,\rbar) \nc
\stp
and  determine generating functional
\st
 Z[J_1,J_2] = Z[0,0] \exp\left(-\frac{1}{2}\int \frac{\dd\omega}{2\pi}\dd r \dd \rbar \; J_{s}(-\omega,r) \left[iG(\omega,r,\rbar) \right]_{ss'} J_{s'}(\omega,\rbar)\right)  \np
\stp

\subsection{Bulk Picture of  Brownian Motion} 
\label{Bulk_brownian}

We will now develop a bulk picture of the Brownian motion. 
Rather than integrating out the entire bulk to determine an effective
action for the boundary point,
we will introduce a stretched horizon and  integrate out only 
the fields which are inside the stretched horizon,
\st
  r_{h} \equiv 1 + \epsilon \np
\stp
The path integral for the fluctuations of the string is
\st
Z = \int \left[\Dt x^{o}_1\,\Db x_1\, \Dt x^{h}_1\right]  \,
\left[\Dt x^{o}_2\,\Db x_2\, \Dt x_2^{h} \right]  \,
\left[\Db x_1^{\epsilon}\, \Db x^{\epsilon}_2 \right] \,  e^{iS_1-iS_2} e^{iS_1^{\epsilon} - iS_2^{\epsilon} }  \np
\stp
As before for the right quadrant for example,  $\Dt x^{o}_1$ is the temporal path
integral of the boundary endpoint of the string, 
$\Db x_1$ indicates the bulk path integral 
of the string,  $\Dt x^{h}$ denotes   the 
temporal path integral of the string point on  the stretched horizon,
and $\Db x^{\epsilon}_1$  is  the bulk path integral inside of the 
stretched horizon.   $S_1$ is the action integrated outside 
of the stretched horizon while $S_1^{\epsilon}$ is the 
action integrated inside the stretched horizon. 

In a Gaussian approximation the integrals over the bulk coordinates
inside the stretched horizon can be done. The result is a field
independent determinant times
the exponential of the action 
evaluated with the classical
solution which passes through  $x_1^{h}(\omega)$ and $x_2^h(\omega)$.
This solution should also obey the contour boundary conditions given above.
Substituting a classical solution into the action reduces to a 
boundary term  at the horizon which gives an 
effective action for the horizon dynamics
\st
S_{\rm eff}^{h} =  -\frac{\T(r_h)}{2} \int_{r^h_1}   \!
x^{\epsilon}_1(-\omega,r) \, \partial_r x^{\epsilon}_1(\omega,r) \, \frac{\dd \omega}{2\pi} 
+ 
\frac{\T(r_h)}{2} \int_{r^h_2} \! x^{\epsilon}_2(-\omega,r) \, \partial_r  x^{\epsilon}_2(\omega,r) \, \frac{\dd \omega}{2\pi} \np
\stp

Going through the same procedure outlined above 
we have the  same structure  for the effective horizon action
\st
\label{horizon_eff}
iS_{\rm eff}^{h} = -i\int \frac{\dd \omega}{2\pi} \, 
x^{h}_a(-\omega) \left[ G_{R}^{h}(\omega) \right] x^h_r(\omega)  - \frac{1}{2} \int \frac{\dd \omega}{2\pi} x_{a}^{h}(-\omega) \left[ -(1 + 2n) \Im G_{R}^h(\omega) \right] x_a^{h}(\omega) \nc
\stp
but the retarded correlator is to be evaluated and normalized at the horizon.
Using the near horizon behavior of the retarded solution 
$F_{\omega}(r) \sim \,(1-1/r^4)^{-i\omega/(4\pi T)}$ and $\T(r)$ from
\Eq{Tofr} we find
\begin{align}
G_{R}^{h}(\omega) =& \left.  \frac{\T(r)}{|F(\omega,r)|^2} 
 F_{-\omega}(r) \partial_r F_{\omega}(r)  \right|_{r=r_h} \nc \nonumber\\
                  =& -i \omega \etav \np
\end{align}

The full partition function is now
\st
\label{Zh2}
Z = \int \left[\Dt x^{o}_1\,\Db x_1\, \Dt x^{h}_1\right]  \,
\left[\Dt x^{o}_2\,\Db x_2\, \Dt x_2^{h} \right]  \,
  e^{iS_1-iS_2} e^{iS^{h}_{\rm eff} }  \np
\stp
Now in \Eq{Zh2} we have the two halves of the Kruskal plane 
coupled by the effective action of the horizon. 
We will show that the effect of this coupling is  to
give rise to  thermal noise in the bulk.
We rewrite the action in the $ra$ basis\footnote{
The label $r$ for retarded should not be confused with the radial 
coordinate.}
\st
 x_{r}(\omega,r) \equiv \frac{x_1(r,\omega) + x_2(r,\omega) }{2} \nc \qquad 
 x_{a}(\omega,r) \equiv x_1(r,\omega) - x_2(r,\omega)  \nc
\stp
and  the action in  the bulk is
\st
iS_1 - i S_2  = -i\int \frac{\dd\omega}{2\pi} \dd r \,  \left[ \T(r) \partial_r x_a(-\omega,r) \, \partial_r x_r(\omega,r) - \frac{\m \omega^2 x_r(\omega,r) x_a(-\omega,r)}{f} \right] \np
\stp
We also follow what is now a standard procedure by introducing 
a horizon noise 
\st
e^{-\frac{1}{2}\int \frac{\dd\omega}{2\pi} x_a^h(-\omega) \, \left[
(1+2n) \omega \etav \right] \, x_a^h(\omega)  } 
=\int D\xi^{h} e^{+i\int  \xi^{h}(\omega) x_{a}^{h}(-\omega) } 
e^{ -\frac{1}{2} \int \frac{\dd \omega}{2\pi} \frac{\xi^h(-\omega) \xi^h(\omega) }{\left[(1+2n)\omega \etav \right] } }  \nc
\stp
with the associated statistics
\st
        \llangle \xi^h(-\omega) \xi^h(\omega) \rrangle  = (1 +2 n) \omega \eta \np
\stp

We now integrate by parts and obtain two boundary terms, one
 terminating at the radius of the string endpoint $r_m$, 
and one terminating at the stretched horizon $r_h=1+\epsilon$
\begin{align}
iS_1 - i S_2 + iS^{h}_{\rm eff}  =& -i \int_{r_m} \frac{\dd\omega}{2\pi} \,  
x_a^o(-\omega) \left[   \T(r_m) \partial_r x_r( \omega,r) \right]   \nonumber \\
               &  - i \int_{r_h} \frac{\dd\omega}{2\pi}  \, x_a^{h}(-\omega) \left[ -\T(r_h) \partial_r x_r(\omega,r)  -  i \omega \etav x_r^h(\omega) - \xi^{h}(\omega)  \right]   \nonumber \\
              & 
- i \int \frac{\dd \omega}{2\pi} \dd r \, x_{a}(-\omega,r)  \,
  \left[ -\partial_r \big( \T(r) \partial_r  x_r(\omega,r) \big)     
- \frac{m \omega^2 \,x(\omega,r)}{f}\right ] \np
\end{align}
Finally integrating over  $x_{a}^o(-\omega)$, $x_{a}(-\omega,r)$, and $x_a^h(-\omega)$
in \Eq{Zh2} we are left with a set of stochastic equations  as
in the simple example given in \Sect{contour_langevin}. We will record these in the next section.

\subsection{Analysis and Discussion}
\label{analysis}

The preceding analysis leads to three equations
\begin{enumerate}
\item The boundary endpoint of the string obeys the 
deterministic equation
\st
\label{boundary}
   -\T(r_m) \partial_r x_r(\omega,r) = 0 \np
\stp
This is simply the Neumann boundary condition for the free end of the 
string.

\item The body of the string obeys the  bulk equations of motion
\st
\label{bulk}
  \left[ \partial_r \big( \T(r) \partial_r x_r(\omega,r) \big)  + \frac{m \omega^2}{f} x_r(\omega,r) \right ] = 0 \np
\stp

\item Finally, the horizon endpoint obeys the 
stochastic equation of motion
\st
\label{Stochastic}
 \T(r_h) \partial_r x_r(\omega,r) + \xi^{h}(\omega) = -i\omega \etav x^{h}_r(\omega) \nc \qquad  \llangle \xi^h(-\omega) \xi^h(\omega) \rrangle = \etav \omega (1 + 2n) \nc
\stp
\end{enumerate}
with $n(\omega) = 1/(\exp(\omega/T) - 1) $.
Here we  remind that 
$\T(r) = (\sqrt{\lambda}\pi^2 T^3/2) fr^4$ is the
local tension in the string.  The meaning of this equation is 
that the motion of  the horizon endpoint $x_{r}^h(t)$ is overdamped. The
resistance  $-\eta \dot x^h$
exactly balances the applied forces  which in this case are
the  pulling due to the string outside the horizon  
$\T(r_h) \partial_r x_r$
and the random force  $\xi^h(\omega)$ which comes from integrating out modes 
behind the horizon.

We now have determined a stochastic equation of motion  for the 
endpoint of the string on the stretched horizon.
These stochastic fluctuations
are transmitted by the dynamics of the string to the boundary. 
We should show that the boundary endpoint of the string obeys the expected 
equation of motion in its most general form, \Eq{GeneralizedLangevin}. 
We should further show that the symmetrized two point functions
obey the expected  dynamics previously computed using 
the Kruskal extension of AdS/CFT
\begin{align}
 \llangle \Delta x_r(-\omega, r) \Delta x_r(\omega,\bar r)  \rrangle =& -(1 + 2n) \Im G_{\ret}(\omega,r,\bar r) \nc \nonumber  \\
 =& (1 + 2n) \frac{\Im F_\omega(r) \Im_\omega F(\rbar) }{-\Im G_{R}(\omega) } \np
\end{align}

First consider the average motion of  the string. Averaging
over the noise we find the average string coordinates 
obey the usual equations of motion together with retarded boundary
conditions  
\st
\etav \left[  4\pi T (1-1/r) \partial_r \llangle x(\omega,r)\rrangle    + i\omega  \llangle x^{h} \rrangle  \right] = 0  \nc
\stp
where we have used the definition for $\T(r)$. 
This equation together  with the fact that $\llangle x_r \rrangle$ is a solution
says that the average obeys the retarded boundary conditions, i.e. behaves 
as $(1-1/r)^{-i\omega/4\pi T}$.

Next consider the behavior near the boundary of AdS $r\rightarrow r_m$.
Here the equations of motion guarantee that the solution is
a superposition of the non-normalizable mode and the  normalizable mode 
which we can choose to be $F_{\omega}(r)$ and $\Im F_{\omega}(r)$  respectively
\st
\label{form_boundary}
 x(\omega,r)= x_o(\omega) F_{\omega}(r) + \xi^{o}(\omega) \left[ \frac{ \Im F_\omega(r)  }{-\Im G_{R}(\omega) } \right] \np
\stp
We also have specified the boundary value of the non-normalizable mode $x_o(\omega)$ and have recognized that $\xi^{o}(\omega)$ must be a stochastic
variable since the retarded solution reproduces the average motion. 
We have chosen to divide $\Im F_\omega(r)$ by $-\Im G_R(\omega)$ 
so that the $\xi^{o}(\omega)$  has the interpretation as the random force 
on the AdS boundary  as we will show now.

Plugging this functional form \Eq{form_boundary} into the Neumann boundary  conditions
\st
\left. \T(r) \partial_r x_r \right|_{r=r_m} =0 \nc
\stp
yields  the expected  equation of motion for the endpoint
\st
   \left[-M_{Q} \omega^2  + G_{R}(\omega) \right] x_o(\omega) = \xi^o (\omega)  \np
\stp
Our task now is to  show that $\llangle \xi^o(-\omega) \xi^o(\omega)  \rrangle$ obeys
the fluctuation-dissipation relation
\st
 \llangle \xi^{o}(-\omega) \xi^{o}(\omega) \rrangle = -(1 + 2n) \Im G_R(\omega) \np
\stp

Returning to the horizon we have the stochastic equation of 
motion
\st
\label{horizon_motion}
 \T(r_h) \partial_r x_r(\omega,r) + \xi^{h}(\omega) = -i\omega  \etav x_r^h(\omega) \np 
\stp 
Substituting  \Eq{form_boundary}  into this equation we  solve 
for $\xi^{o}(\omega)$ in terms of $\xi^h(\omega)$. Our task is simplified by first  recognizing the origin
of the first term on the right hand side
\st
\label{horizon_origin}
     -i\omega \etav = \T(r_h) \frac{F_{-\omega}(r_h) \partial_r F_{\omega}(r_h) }{|F_{\omega}(r_h)|^2 } \nc
\stp 
so that  the equation  for $\xi^{o}$  resulting from
\Eq{horizon_motion} is
\st
\frac{\xi^{o}(\omega)}{-\Im G_R(\omega) } \T(r_h) \left[ F_{\omega}(r_h) \partial_r \Im F_{\omega} - \Im F(r_h) \partial_r F_{\omega}(r_h) \right]  + F_{\omega}(r_h) \xi^{h}(\omega)  = 0  \np
\stp
The term in square brackets on the left hand side is 
the Wronskian  of the two solutions and when multiplied 
by $\T(r)$ equals $+\Im G_{R}(\omega)$ 
and we find
\st
\label{horizon_bndry_relation}
  \xi^{o}(\omega) =  F_{\omega}(r_h) \xi^{h}(\omega)    \np
\stp 
Averaging  according to the horizon statistics in \Eq{Stochastic}, recognizing its origin  \Eq{horizon_origin},  and again using the 
Wronskian relation \Eq{ImGrindep}, we determine the statistics
of $\xi^{o}$
\st
\label{boundary_stat}
\llangle \xi^{o}(-\omega) \xi^{o}(\omega) \rrangle  = -(1 + 2n) \Im G_{R}(\omega) \np
\stp
Thus random force in the boundary theory obeys the expected statistics of
the fluctuation-dissipation theorem.

Finally we would like to compute the bulk two point functions 
\st
 \llangle \Delta  x_r(-\omega, r) \Delta x_r(\omega, \bar{r}) \rrangle \nc
\stp
where $\Delta x_r(\omega,r) $ 
denotes the deviation from the  behavior in 
the bulk due to the motion on the boundary 
$\Delta x_r \equiv x_r(\omega,r)  - x^{o}_r(\omega) F_{\omega}(r)$.  
This is straightforward using the decomposition in \Eq{form_boundary}, and 
the boundary statistics in \Eq{boundary_stat}; the result is 
\begin{align}
\label{gsymbulk}
\llangle \Delta  x_r(-\omega, r) \Delta x_r(\omega, \rbar) \rrangle =& 
(1 + 2n) \frac{\Im F_{\omega}(r) \Im F_\omega( \bar r)  }{- \Im G_R(\omega) } \nc \\
 =& -(1 + 2n) \Im G_\ret(\omega, r,\rbar) \np
\end{align}
This result is naturally  the  same as computed previously using 
the contour correlation function in AdS/CFT. It is also neatly consistent 
with the bulk fluctuation-dissipation theorem. 

\section{Summary and the Physical Picture}
\label{Summary}

\subsection{Summary}
\begin{figure}
\begin{center}
\includegraphics[height=2.0in]{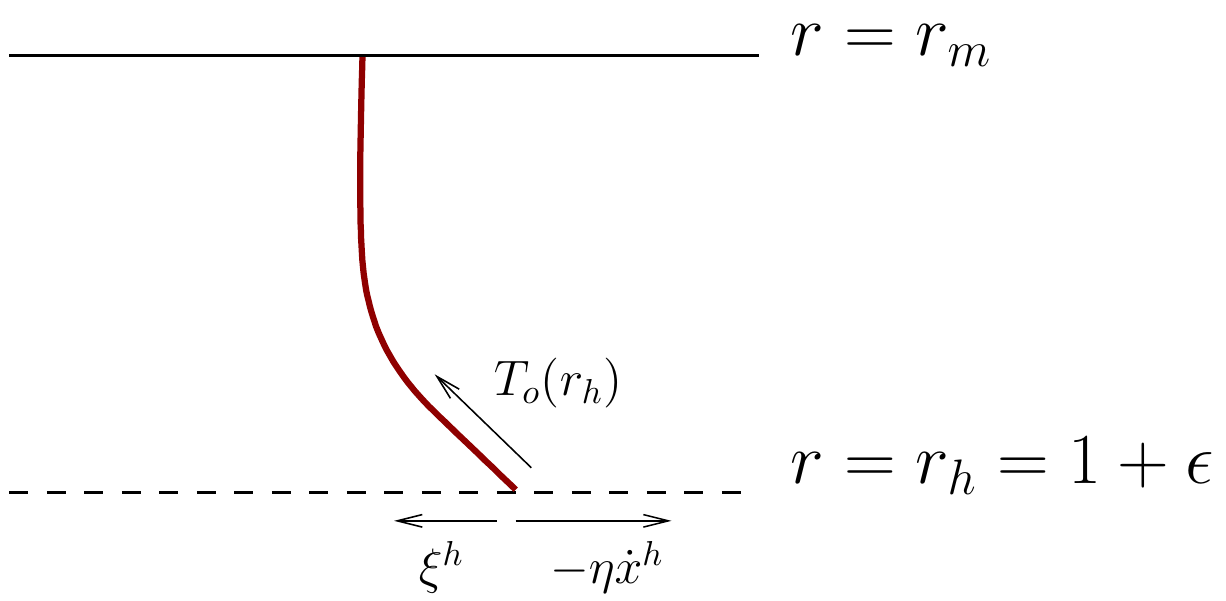}
\end{center}
\caption{
Balance of forces on the stretched horizon. The resistive force 
$-\eta \dot{x}^h$
precisely balances the random force $\xi^h$ and the tension $\T$ 
leading to overdamped motion. }
\label{horizon}
\end{figure}

In the previous section we have shown  that 
a stochastic boundary condition emerges 
on the stretched horizon after integrating out the 
fluctuations inside this surface  (see \Eq{Stochastic})\footnote{In 
this section the retarded $r$ label (as in $x_r$)  is understood.}
\st
\label{horizon_bound}
\T(r_h) \partial_r x(t,r_h)  + \xi^{h}(t) = \etav \,\dot x^{h}(t)   \qquad \llangle \xi^h(t) \xi^h(t') \rrangle  = G_{\rm sym}^h(t-t') \np
\stp
Here $\etav$ is the drag 
of the horizon (or the late time drag of the quark),  $\xi^{h}$ 
is the random force on the horizon endpoint, and $\xi^h(\omega)$ 
obeys the {\it horizon}
fluctuation dissipation theorem
\st
 G_{\rm sym}^{h} (\omega) =  + (1 + 2n) \omega \etav \np
\stp
This equation
has a simple physical interpretation illustrated in 
\Fig{horizon}. 
The motion 
of the horizon $x^{h}(t)$ is overdamped, $i.e.$ the resistive force $-\etav \dot x^h$ exactly
balances the string force $\T(r_h) \partial_r x(t,r_h)$  
and the random horizon  force $\xi^{h}$. There is no transverse acceleration.

This stochastic force on the horizon when transmitted 
to the boundary
leads to an equation of motion  for the endpoint of the string
\st
 M_{Q} \frac{d^2x^o}{dt^2} + \int^{t} G_{R}(t-t') x^o(t')  = \xi^{o}(t) \nc  \qquad \llangle \xi^{o}(t) \xi^{o}(t') \rrangle  = G_{\rm sym}(t,t')  \np
\stp
which is the expected generalized Langevin equation in \Eq{GeneralizedLangevinT}.
Here $G_{\rm sym}(\omega)$ obeys {\it boundary} 
the fluctuation dissipation relation
\st
  G_{\rm sym}(\omega)  =  -(1 + 2n) \Im G_{R}(\omega) \nc
\stp
and $G_{R}(\omega)$ is the usual retarded force-force propagator computed 
using AdS/CFT. 
The random force  
on the boundary $\xi^o(t)$ is directly related to the random force on the 
horizon $\xi^h$.  Denoting $F(t,r)$ the usual retarded boundary to bulk propagator,
$i.e.$ $F_{\omega}(r_m) =1$ on the boundary and $F_{\omega} \sim (1-1/r)^{-i\omega/4\pi T} $ near the horizon, the relation between the horizon and boundary 
stochastic forces is (see \Eq{horizon_bndry_relation}) 
\st
   \xi^{o}(t)  =  \int^t \dd t'\, F(t - t', r_h)\, \xi^{h}(t') \nc
\stp
It takes a time of order $ \sim 1/(\pi T) \log(1/\epsilon)$ 
for the noise from the horizon  to reach the tip of the string. $\epsilon$
should be considered small but not exponentially small so this
time scale is really $1/\pi T$.

We also have a picture of the fluctuations in the bulk. The coordinate
of the string  in the bulk are given by two
pieces (see \Eq{form_boundary}) 
\st
   x(t,r) =  \int^{t} \dd t' F(t- t',r) x^{o}(t') + \Delta x(t,r)  \nc
\stp
which reflect the retarded response to the boundary motion $x^{o}(t')$ and a deviation.
The deviation $\Delta x(t,r)$ is a random variable  obeying  the statistics (see \Eq{gsymbulk})
\st
  \llangle  \Delta x(t,r) \Delta x(t', r') \rrangle = G_{\rm sym}(t-t',r,r') \nc
\stp
where $G_{\rm sym}(t-t',r,r')$ is the symmetrized bulk to bulk correlator. 
This correlator was computed using  the Kruskal extension of the 
AdS/CFT and is related to the imaginary part of the retarded bulk to bulk 
correlator  according to a {\it bulk}  fluctuation dissipation
theorem
\st
    G_{\rm sym}(\omega, r, r') = - (1 + 2n) \, \Im G_{\ret}(\omega,r,r') = (1 + 2n) \,\frac{\Im F_\omega(r) \Im F_\omega(r') }{-\Im G_R(\omega) } \np
\stp
The explicit form for the fluctuation amplitude 
in frequency space is
\st
\label{fluctamp}
     \Delta x(\omega,r) = \xi^{o}(\omega)\, \left[ \frac{\Im F_{\omega}(r) }{-\Im G_R(\omega) }\right] \nc
\stp
where $\xi^{o}(\omega)$ is the boundary force.

\subsection{The Physical Picture}
Here we would like to consider the small frequency limit where an explicit 
analytic form for the retarded function $F_{\omega}$ is known and has
a simple physical interpretation 
in terms of trailing strings.
A quark in equilibrium moves quite slowly 
\st
   v_{\rm th} \sim \sqrt{\frac{T}{M_{\rm kin}}}  \sim \frac{1}{\lambda^{1/4}} \frac{1}{\sqrt{r_m}}\,  \nc
\stp
but it takes a long time $\tau_R$ for this heavy quark to randomize its velocity
\st
  \tau_R \sim  \frac{M_{\rm kin}}{\eta } \sim \frac{r_m}{T} \np
\stp
(This can be seen by examining the Langevin equations and neglecting the 
noise.)   The distance the quark moves  over this 
relaxation time $x_R$ is  
\st
    x_{R} \sim  v_{\rm th} \tau_R \sim  \frac{1}{\lambda^{1/4}\, T} \sqrt{r_m} 
\np
\stp
The dynamics that is observed depends on how the spatial 
$x_{\rm obs}$ and temporal resolution scales $\tau_{obs}$ 
of the measurement compare to these scales $x_R$ and $\tau_R$. 

First consider the  time period  over which
 quarks moves with nearly constant velocity $v$
\st
  \frac{1}{T} \ll  \tau_{\rm obs} \ll \tau_R \np
\stp
For a quark moving slowly 
on the boundary with constant velocity $v$  it will trail behind
it a trailing string at least on average
\st
\llangle x(t,r ) \rrangle = x_o(t)  
+ v \xts(r)  \nc
\stp
with
\st
  \xts(r) = \frac{1}{2\pi T} \left[ \tan^{-1}(z) - \tanh^{-1}(z) \right]  \nc
\stp
and $z\equiv1/r$.
Here we have used \Eq{form_boundary} for retarded 
response to the boundary motion
and the explicit form of $F_{\omega}(r)$ 
at small frequency, \Eq{trailing_string}. 
The term  $\xts$
is the ``trailing string" solution of \Ref{HKKKY,Gubser}.  
The distance between the head of the quark $x^{o}(t)$ and 
the average body of string is  of order
\st
 v\xts  \sim \frac{v_{\rm th}}{T} \sim \frac{1}{\lambda^{1/4} T} \frac{1}{\sqrt{r_m} } \np 
\stp

\begin{figure}
\begin{center}
\includegraphics[height=2.2in]{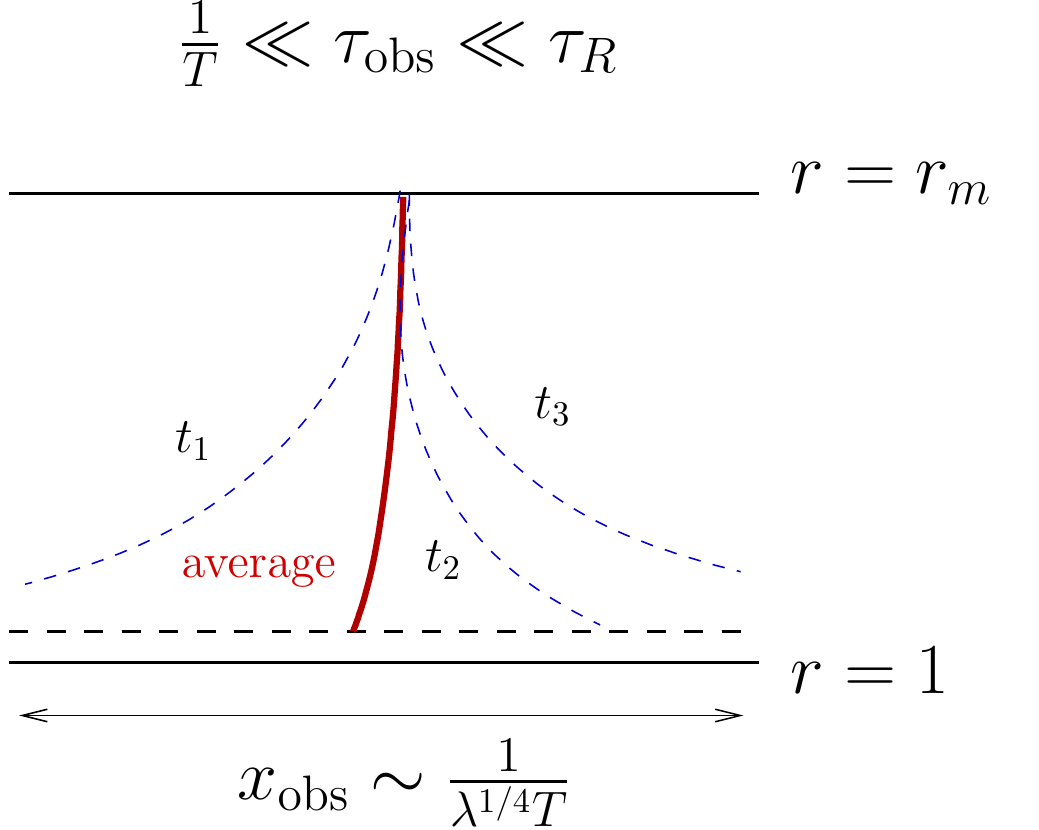}
\includegraphics[height=2.2in]{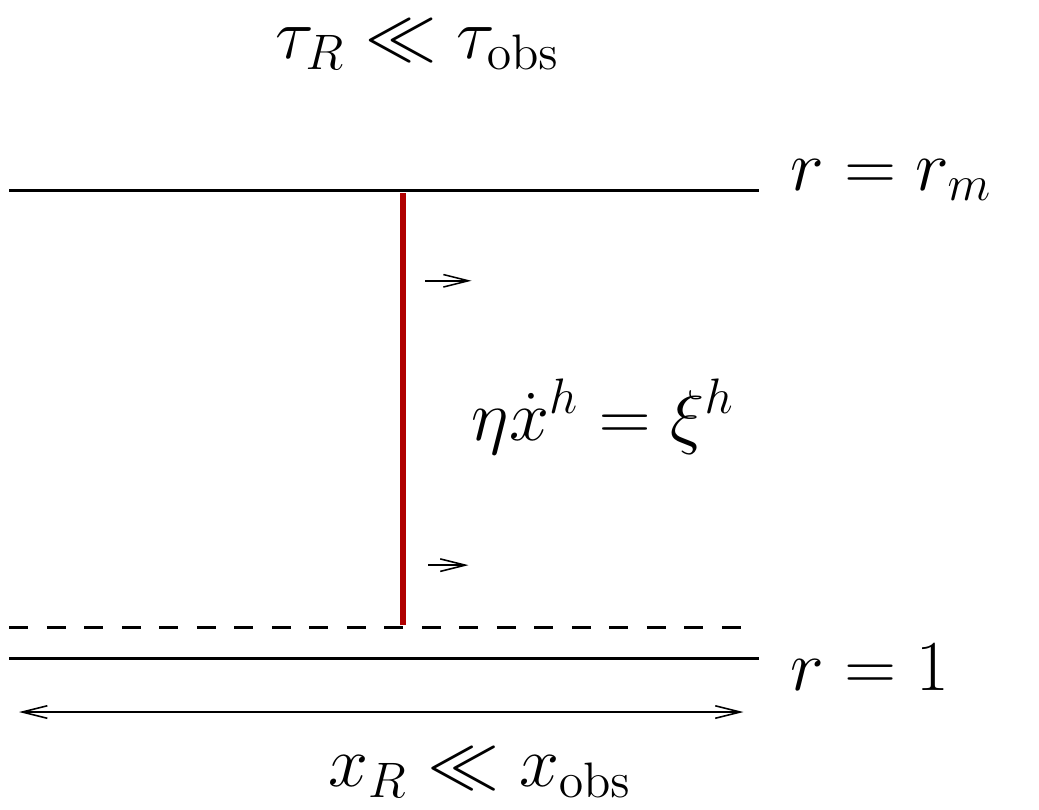}
\caption{(a) The physical picture that emerges 
when observing the quark on relatively short time 
scales $1/T \ll \tau_{\rm obs} \ll \tau_R$. Here 
we show three subsequent time steps, $t_1, t_2,t_3$; at 
each time step the string fluctuates to a new ``trailing string"
giving  rise to a random force on the boundary. The average
trailing string is perceived as a drag.
(b) The physical picture that  emerges on very long time
and spatial scales. The horizon diffuses and the string is 
brought along. 
\label{trailing_string_fig}
}
\end{center}
\end{figure}

This only gives 
the average behavior of the string.  
In general there is 
an additional random component  
which in the small frequency limit is white noise; using \Eq{fluctamp} 
we have
\st
\label{flip_flop_string}
 \Delta x \equiv x(t,r) - \llangle x(t,r) \rrangle =\frac{-1}{\etav} \xi^{o}(t) \, 
\xts(r)  \nc \qquad  \llangle \xi^{o}(t) \xi^{o}(t') \rrangle = 2 T \etav \delta(t-t') \np
\stp
Thus we see that around this average trailing string there is a stochastic ensemble of trailing strings  which flip-flop around the head of the quark.
This is illustrated above in \Fig{trailing_string_fig}.  To estimate the amplitude
of this stochastic process, let us imagine implementing this process on the 
computer where one would take time step $\Delta t$ and then the width of the Gaussian 
process would be $ \llangle \xi^{o} \xi^{o} \rrangle = 2 T \etav/\Delta t$. 
Taking the width $\Delta t$ to be of order the memory time scale $1/T$,
we estimate  that the string fluctuates around  the average trailing string
by an amount 
\st
\sqrt{(\Delta x)^2 }  \sim  \frac{1}{\sqrt{\eta} } \sim \frac{1}{\lambda^{1/4} T} \nc
\stp
This is larger than the average deviation $v \xts$ from the endpoint  since 
it is not suppressed by  $1/\sqrt{r_m}$. Thus the average
trailing string  is nearly straight and the noise consists of a 
flip flopping trailing string solution. Notice the minus 
appearing in  front of the \Eq{flip_flop_string}. This is physically 
correct. When the random force on the boundary quark is positive, the string
is out in front of the quark. When the random force is negative, the string
trails behind the quark.  

Now let us consider the ultimate long time limit where the 
time scales and spatial scales 
on which we are observing the quark
are large compared to the relaxation time $\tau_R$  and relaxation length $x_R$
\st
   \tau_{\rm obs} \gg \tau_R  \nc
\qquad
 x_{\rm obs} \gg v_{\rm th} \tau_R \gsim \frac{1}{\lambda^{1/4} } \frac{1}{T} \sqrt{r_m} \np
\stp
In this limit one can drop the influence 
of the bulk on the horizon dynamics, $i.e.$ discard
the  $T_{o} \partial_r x$  in \Eq{horizon_bound}  since
it is averaged over many different boundary velocities 
over the observation time $\tau_{\rm obs} \gg \tau_R$. 
The string
in bulk is perfectly straight on the spatial scales we are considering.
Thus the equation of motion obeyed  by the horizon is
\st
   \frac{\xi^h}{\eta}  =  \frac{dx^h}{dt}   \np
\stp
This is the overdamped diffusive limit discussed  in \Sect{basic}.  
The result is that the horizon endpoint and  the boundary 
endpoint diffuse in  lockstep  according to the expected rate 
\st
\label{horizon_diffusion}
  \llangle [x^{h}(t)]^2 \rrangle = \frac{2 T}{\eta} t  \np
\stp
\newline


In summary, we have exhibited the full structure of the thermal noise on a 
fluctuating string in AdS/CFT. It is hoped that this will  
lay the groundwork to study the fluctuations of gravitons and 
other fields using the correspondence. The challenge now is to
use the real time formalism in a truly out of equilibrium setting
such as studied in \Ref{Chesler:2008hg}. 
\newline

{ \noindent {\bf Acknowledgments.}  We gratefully acknowledge 
discussions with Jorge Casalderrey-Solana, Edward Shuryak, 
Ismael Zahed, Elli Pomoni, and Keun-Young Kim. D.T.S. is supported, in part, by
DOE grant DE-FG02-00ER41132.
D.T. is supported by the U.S. Department of Energy under an OJI
grant DE-FG02-08ER41540 and as a Sloan Fellow. 
}
\newline

{ \noindent {\bf Note Added. } In the final two weeks of preparation,
a preprint appeared \cite{deBoer:2008gu} which addressed some of the aspects
studied here. In particular the stochastic  boundary condition given 
in \Eq{horizon_bound}  was determined 
and the basic conclusion that the horizon 
diffuses according to \Eq{horizon_diffusion} was also reached in this work.
It is interesting that the derivation of the 
stochastic boundary condition 
in \Ref{deBoer:2008gu} differs substantially from the 
presentation given here. Presumably by marrying the two derivations
a deeper understanding of Hawking radiation will be reached.
\appendix

\section{Small Fluctuations and the Trailing String }
\label{fluctuations}

The purpose of this appendix is to establish notation and 
to collect prior results.
The  Nambu-Goto action  is 
\st
S = -\frac{1}{2\pi \ls2} \int \dd\tau \dd\sigma \, \sqrt{-{\rm det}\, h_{ab}} \np 
\stp
For small fluctuations we parameterize the string as
\st
  (\tau, \sigma) \mapsto  (t = \tau, r=\sigma, x=x(t,r) ) \nc
\stp
determine the induced metric,
\begin{subequations}
\bg
h_{tt} &=& \left[-\f r^2 + r^2 \dot x^2\right] L^2(\pi T)^2 \nc \\
h_{rr} &=& \left[\frac{1}{\f  r^2} +  (\pi T)^2 r^2 (x')^2\right] L^2 \nc \\
h_{tr} &=& \left[ (\dot x) x' r^2 (\pi T)^2 \right] L^2 \nc
\nd
\end{subequations}
and write the action for small fluctuations as
\st
\label{action_append}
S = -\frac{(\pi T) L^2}{2\pi \ls2} \int \dd t \dd r \left[1 + \frac{1}{2}(\pi T)^2{\f}r^4(x')^2 
 - \frac{1}{2} \frac{{\dot x}^2}{\f} \right] \np
\stp  
In the body of the text we write this as
\st
S = -\int \dd t \dd r \left[\m + \frac{1}{2}\T(r)(\partial_r x)^2 
 - \frac{\m}{2\f} (\partial_t x)^2  \right] \nc
\stp
where the local tension is
\st
  \T(r) = \frac{(\pi T)^3 L^2}{2\pi\ls2}\,  fr^4   =  (\sqrt{\lambda} \pi^2 T^3/2) \,  fr^4 \nc
\stp 
and the mass  per unit $r$
\st
   \m = \frac{(\pi T) L^2}{2\pi \ls2} = \frac{\sqrt{\lambda} T}{2} \np
\stp
Then the equation of motion is 
\st
\label{eom}
  \frac{\wn^2}{f}x + \partial_r ( f r^4 \partial_r x) = 0 \np
\stp
where,  as is customary, we have defined $\wn = \omega/(\pi T)$.

This is a second order differential equation and there are two solutions.  
Near the horizon $r\rightarrow 1$ the solutions are either infalling (-)  or outgoing (+)
\st
 \left(1 - \frac{1}{r^4}\right)^{\mp\frac{i\wn}{4} }  \np
\stp 
The solutions  near the boundary
$r\rightarrow r_m$  consist of a normalizable and a non-normalizable mode.
The retarded  solution $F_{\omega}(r)$ is the solution  
which approaches one near the boundary and is infalling at the horizon.
More specifically near the boundary  $F_{\omega}$ behaves as
\st
F_{\omega}(r)  = \Big(1  + \frac{\wn^2}{2 r^2} +  \ldots\Big)  
-\frac{B(\omega)}{3r^3} \Big( 1   + \ldots \Big) \nc
\stp
where the ellipses denote  terms suppressed by additional powers of $1/r$.
In evaluating the motion of the quark we often evaluate the combination
\st
 \lim_{r\rightarrow r_m} \T(r) F_{-\omega} \partial_r F_{\omega}(r) = -M_{Q}^{o} \omega^2 +  G_{R}(\omega) \nc
\stp
where $G_{R}(\omega)=(\sqrt{\lambda}\pi^2 T^3/2) B(\omega)$ is the retarded force-force correlator.
The first term $M_{Q}^{o}\omega^2$ comes from the ``divergent" $\w^2/2r^2$ 
term of the real part of the retarded Green function and 
we have identified
\bg
 M^{o}_{Q} =  \frac{L^2 (\pi T) r_m}{2\pi \ls2}  = \frac{ \bar r_{m} }{2\pi\ls2} \nc
\nd
as the  zero-temperature mass of the quark  where 
$\bar r_m$ refers to the ``canonical" coordinates in \Eq{canonical}. 
That this is the zero temperature mass
can also be seen from the leading term of \Eq{action_append}.

In general \Eq{eom} can not be solved exactly. However we can set up a perturbation
expansion at small frequencies,
\st
F_{\omega}(r) =(1 - 1/r^4)^{-i\wn/4} \left[1 - i\wn F_{\omega}^{(1)}(r) - \wn^2 F_{\omega}^{(2)}(r) + \ldots \right] \np
\stp
Substituting this ansatz into the equation of motion we 
end up with a hierarchy of differential equations. We solve these equations order by order in $\omega$  by demanding that the solution behaves as
\st
 F_{\omega}(r) = \left(1 - \frac{1}{r^4}\right)^{-i\wn/4} \times \left( \mbox{regular function at the horizon} \right)  \np
\stp
We find
\st
 F_{\omega}^{(1)}(r) = 
   -\frac{1}{2} \ln  \left( z+1 \right)  -\frac{1}{4} \ln  \left( {z}^{2}+1
 \right) +\frac{1}{2} \arctan \left(z \right) \nc 
\stp
where here and below $z \equiv 1/r$. 
At quadratic order  we have 
\begin{align}
F_{\omega}^{(2)}(r) =&
 \int_0^{\frac{1}{r}}dz {\frac {z \left( 1-z \right)  \left[ z\ln  \left( {z}^{2}+1 \right) -2\,z\arctan
 \left( z \right) +2\,\ln  \left( z+1 \right) z-4 \right] }{4(1-{z}^{4})}} \np 
\end{align}

This solution has a simple physical  interpretation. The 
leading order solution is multiplied by the 
position of the tip string $x_o(\omega)$. Provided we are not
exponentially close to the horizon we can expand the leading  
$(1-z^4)^{-i\wn/4}$ factor. This yields the  solution 
\st
\label{expansion}
   x_o(\omega,r) \equiv x_o(\omega)F_{\omega}(r) = x_o(\omega) +
v(\omega) \xts + a(\omega)
\Delta x_{a} \nc
\stp
where $v(\omega) = -i\omega x_o(\omega)$ is the velocity of the
endpoint, and  $a(\omega) = (-i\omega)^2 x_o(\omega)$ is
the acceleration of the endpoint. Further  we have defined
\begin{subequations}
\begin{align}
 \xts(z) =& \frac{1}{2\pi T} \left[\tan^{-1}(z) -  \tanh^{-1} (z) \right]  \nc \\
 \Delta x_{a}(z) =& 
\frac{1}{(\pi T)^2} \left[\frac{1}{32}\log^{2}(1-z^4) + \frac{1}{4} \log(1-z^4) F_{\omega}^{(1)} + F_{\omega}^{(2)} \right] \np
\end{align}
\end{subequations}

The leading term is given by $\xts$ which is simply the 
``trailing string" solutions of \Ref{HKKKY,Gubser}. Naturally to 
leading order in the frequency of the sinusoidal oscillation one 
simply recovers that the string trails behind the head of the quark
with the expected form.

The subleading term is described by the acceleration $\Delta x_{a}$.
A graph  
of this function is
given in \Fig{afigure}  and has 
a simple interpretation.
Consider sinusoidal oscillations:
when the head of the quark is moving forward and undergoing negative acceleration, 
the body of the string travels ahead of the trailing string solution due
to inertia. Thus $-\Delta x_a$ should be positive as $r\rightarrow \infty$ 
reflecting the fact that the displacement is $180^{o}$ out of phase with the 
acceleration. This inertial effect is indicated by the $1/2r^2$ curve 
in \Fig{afigure}. 
The dynamics close to the horizon stems from expanding out the 
leading $(1-z^4)^{-i\wn/4}$ factor. The $(1-z^4)^{-i\wn/4}$ behavior near  the 
horizon has the interpretation that the string endpoint on 
the stretched horizon is overdamped.
\begin{figure}
\includegraphics[height=3.2in]{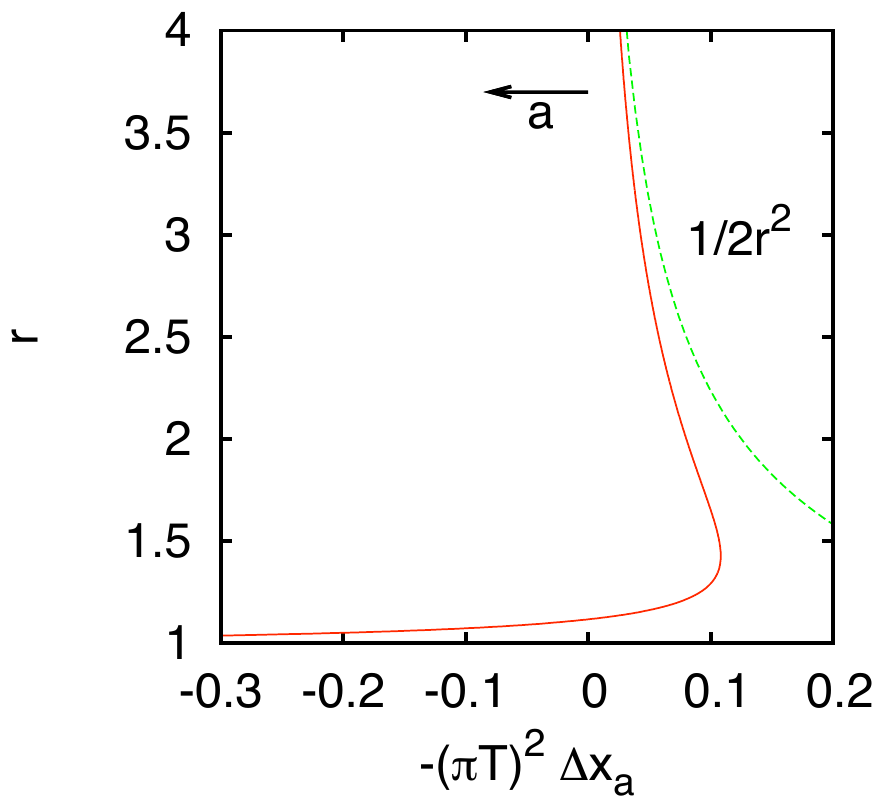}
\caption{
The deviation of the string position 
from the trailing string solution (in units of $|a(t)|/(\pi T)^2$)
during slow sinusoidal acceleration of the 
boundary endpoint. Here the acceleration is negative and the quark is 
moving forward. More specifically we are plotting 
$-(\pi T)^2 \Delta x_a$ given in the text.
\label{afigure}
}
\end{figure}

Using  the solutions given above, we can expand these functions close to 
the boundary 
\st
 F_{\omega}(r) = \Big(1 +  \frac{\wn^2}{2 r^2} + \ldots \Big) + \frac{(i\wn - \wn^2) }{3r^3} \Big( 1 + \ldots \Big) \nc
\stp
and determine the retarded force-force correlator 
\begin{align}
\T(r) F_{-\omega} \partial_r F_{\omega} &= -M_{Q}^{o}\omega^2 +  G_{R}(\omega) 
\nc  \\
&= -M_{Q}^{o} \omega^2 -i\omega \etav  - \omega^2 \Delta M \np
\end{align}
Here we have defined the transport coefficient 
\st
\etav = \frac{1}{2}\sqrt{\lambda} \pi T^2 \nc
\stp
first computed in \Ref{HKKKY,Gubser,Jorge} and the in-medium mass shift 
\st
 \Delta M = -\frac{\sqrt{\lambda} T}{2} \nc
\stp
first computed in \Ref{HKKKY}. 
The fact that the mass shift is negative stems from the  overdamped motion
of the horizon endpoint. Also used in the text is the kinetic mass 
\st
 M_{\rm kin} \equiv M_{Q}^{o} + \Delta M \np
\stp

\section{Notation for the Kruskal Plane}
\label{Krusk_notation}

In this appendix we establish notation for the Kruskal variables used in the body of
the text.
We  first define $r_{*}(r)$
\st
  r_{*}(r) = \frac{1}{\pi T} \int^{r} \frac{dr}{f(r) r^2} = \frac{1}{2\pi T} \tan^{-1}( r)  +
\frac{1}{4\pi T}  \ln( r -1) - \frac{1}{4\pi T} \ln( r + 1)  \np
\stp
Then $U$ and $V$  are defined by the relations 
\begin{subequations}
\st
   t = \frac{1}{4\pi T} \log(V) - \frac{1}{4\pi T} \log(-U) \nc  
\stp
\st
   r_{*} = \frac{1}{4\pi T} \log(V) + \frac{1}{4\pi T} \log(-U) \np  
\stp
\end{subequations}
The near horizon behaviors
of $\nu_-\equiv t+r_*$ and $\nu_+\equiv t-r_*$ are
\begin{subequations}
\st
 \nu_{-} \simeq t + \frac{1}{4\pi T} \log(r - 1) \nc \\ 
\stp
\st
 \nu_{+} \simeq t - \frac{1}{4\pi T} \log(r - 1) \np
\stp
\end{subequations}
Also note that $-UV$ is  a simple function of $r$
\bg
\label{UVrelation}
    (-U) V = e^{4r_* \pi T}  =  \frac{r - 1}{r + 1} e^{2\tan^{-1}(r) } \np
\nd
}

\end{document}